\documentclass[a4paper,11pt]{article}

\usepackage{graphicx}
\usepackage[margin=3cm]{geometry}
\usepackage{epstopdf, epsfig}
\usepackage[utf8]{inputenc}
\usepackage{amsmath} 
\usepackage{amsfonts}
\usepackage{amssymb}
\usepackage{esint} 
\usepackage{setspace} 
\usepackage{comment}
\usepackage{appendix}
\usepackage{enumerate} 
\usepackage[shortlabels]{enumitem} 
\usepackage{url} 
\usepackage{float} 
\usepackage{booktabs} 
\usepackage{upgreek}
\usepackage{wrapfig}
\usepackage{multirow} 
\usepackage{mathtools} 
\usepackage[hidelinks]{hyperref} 
\usepackage{bm} 

\usepackage[authoryear]{natbib}

	\DeclarePairedDelimiter\abs{\lvert}{\rvert}%
	\makeatletter
	\let\oldabs\abs
	\def\abs{\@ifstar{\oldabs}{\oldabs*}}
	\makeatother
	
	\newcommand{\medtab}[0]{\hspace{0.5cm}} 

	\newcommand{\unit}[1]{\ensuremath{\mathrm{#1}}} 
	\newcommand{\un}[1]{\ensuremath{\ \mathrm{#1}}} 

	\newcommand{\dif}[0]{\mathrm{d}} 
	\newcommand{\pd}[2]{\frac{\partial #1}{\partial #2}}
	
	\newcommand{\D}[2]{\frac{\mathrm{D} #1}{\mathrm{D} #2}}
	\newcommand{\pds}[2]{\frac{\partial^2 #1}{\partial #2 ^2}}

	\newcommand{\intn}[4]{\int_{#1}^{#2} \! #3 \, \mathrm{d} #4} 
	
	\newcommand{\conv}[0]{\ast}

	\newcommand{\fts}[1]{\mathcal{F}\lbrace{#1}\rbrace} 
	\newcommand{\ftinvs}[1]{\mathcal{F}^{-1}\lbrace{#1}\rbrace} 
	\newcommand{\stepsymbol}[0]{H_s}
	\newcommand{\step}[1]{\stepsymbol({#1})}
	\newcommand{\expp}[1]{\mathrm{e}^{#1}} 
	\newcommand{\comp}[0]{\mathrm{i}} 

	\newcommand{\goto}[0]{\rightarrow}
	\newcommand{\real}[1]{\mathrm{Re}\{{#1}\}}

	\newcommand{\degree}[0]{^{\circ}}

	\newcommand{\lstokes}[0]{l_S} 
	\newcommand{\ks}[0]{k_s}
	\newcommand{\ksp}[0]{k_{sp}}
	\newcommand{\ksa}[0]{k_{sa}}
	\newcommand{\lsa}[0]{l_{sa}}

	\newcommand{\omeganot}[0]{\omega_0}
	
	\newcommand{\fnot}[0]{f_0}
	
	\newcommand{\phaseang}[0]{\phi} 

	\newcommand{\re}[0]{Re} 

	\newcommand{\halfc}[0]{h} 
	\newcommand{\tc}[0]{\delta} 
	\newcommand{\htot}[0]{h_t} 
	
	\newcommand{\diver}[0]{\nabla \cdot}
	\newcommand{\curl}[0]{\nabla \times}

	\newcommand{\stresssym}[0]{\sigma}
	\newcommand{\stress}[1]{\stresssym_{#1}}
	\newcommand{\stresstenf}[1]{\stresssym_{#1}^{f}}
	\newcommand{\stresstens}[1]{\stresssym_{#1}^{s}}
	\newcommand{\strainsym}[0]{\epsilon}
	\newcommand{\strain}[1]{\strainsym_{#1}}

	\newcommand{\disp}[0]{\xi}
	
	\newcommand{\dispi}[1]{\disp_{#1}}

	\newcommand{\us}[0]{u_s}
	\newcommand{\usnot}[0]{u_{s0}}
	\newcommand{\uf}[0]{u_f}
	\newcommand{\ufnot}[0]{u_{f0}}
	
	\newcommand{\uft}[0]{u_{ft}}
	\newcommand{\usurf}[0]{u_{c}}
	
	\newcommand{\usurfnot}[0]{u_{c0}}
	\newcommand{\usurfnotbreve}[0]{\breve{u}_{c0}}
	\newcommand{\tauw}[0]{\tau_w}
	\newcommand{\tauwnot}[0]{\tau_{w0}}
	\newcommand{\tauwt}[0]{\tau_{wt}}
	\newcommand{\tauwr}[0]{\tau_{wr}}
	\newcommand{\tauwrrigid}[0]{\tau_{wr}^\text{rigid}}
	\newcommand{\deltataur}[0]{\Delta\tau_r}
	
	\newcommand{\dilatmod}[0]{\Lambda} 
	\newcommand{\shearmod}[0]{G} 

	\newcommand{\relaxmod}[0]{\psi}

	\newcommand{\dilatmodrelax}[0]{\relaxmod_\dilatmod}
	\newcommand{\shearmodrelax}[0]{\relaxmod_\shearmod} 
	\newcommand{\complexmod}[0]{M}
	\newcommand{\complexmodreal}[0]{\complexmod_1}
	\newcommand{\complexmodimag}[0]{\complexmod_2}

	\newcommand{\lossangle}[0]{\phase{\complexmod}}

	\newcommand{\dissfac}[0]{\eta_{\complexmod}} 
	\newcommand{\phase}[1]{\phi_{#1}} 
	\newcommand{\phiG}[0]{\phase{G}}
	
	\newcommand{\rhos}[0]{\rho_s}
	\newcommand{\cs}[0]{c_s}
	\newcommand{\csp}[0]{c_{sp}}

	\newcommand{\deltarho}[0]{\delta \rhos}

	\newcommand{\rhof}[0]{\rho_f}

	\newcommand{\fxf}[0]{f_x^f}
	
	\newcommand{\ubreve}[0]{\breve{u}}

	\newcommand{\ufnotbreve}[0]{\breve{u}_{f0}}
	\newcommand{\usnotbreve}[0]{\breve{u}_{s0}}
	\newcommand{\zbreve}[0]{\breve{z}}

	\newcommand{\Wo}[0]{\mathrm{Wo}}
	\newcommand{\omegars}[0]{\omega_{rs}} 
	\newcommand{\omegarf}[0]{\omega_{rf}} 
	\newcommand{\rhor}[0]{\rho_r} 
	\newcommand{\fr}[0]{f_r} 
	\newcommand{\hr}[0]{h_r} 
	\newcommand{\intpar}[0]{\chi} 

	\newcommand{\lb}[0]{\left(}
	\newcommand{\rb}[0]{\right)}

\begin{document}

\title{Oscillatory laminar shear flow over a compliant viscoelastic layer on a rigid base}

\author{H.O.G. Benschop$^{*}$\thanks{$^*$Corresponding author. Email: H.O.G.Benschop@tudelft.nl
\vspace{6pt}}
\, and W.-P. Breugem \\
\vspace{1pt} 
\em{Laboratory for Aero and Hydrodynamics, Delft University of Technology, Delft,}\\
\em{The Netherlands}\\
\vspace{6pt}
}

\maketitle

\maketitle

\begin{abstract}
We present an analytical study of oscillatory laminar shear flow over a compliant viscoelastic layer on a rigid base. This problem relates to oscillating blood flow in viscoelastic vessels. The deeper motivation for this study, however, is the possible use of compliant coatings for turbulent drag reduction.
An analytical solution of the fluid and solid velocity is presented, and five dimensionless parameters emerge. The interaction between fluid and solid appears to be determined by a single combined dimensionless parameter, which we call the shear interaction parameter $\intpar$.
The fluid satisfies a no-slip boundary condition when $\abs{\intpar} \goto 0$, which occurs when the solid is heavy, stiff and/or thin. In contrast, the fluid obeys a free-slip boundary condition when $\abs{\intpar} \goto \infty$, which corresponds to a lightweight and/or soft solid.
Three types of resonance modes are identified for an \textit{elastic} solid. Two modes (odd and even) are specific to the solid. The third mode results from the coupling with the fluid. The three modes are less pronounced or even absent for a \textit{viscoelastic} solid.
These findings have a twofold use. First, they help to understand the fluid and solid dynamics when shear coupling is important. Second, the presented analytical solution is very useful for validation of numerical fluid-structure-interaction solvers.
Future work might include the extension of the theory to multiple viscoelastic layers and the dynamic coupling of normal stresses.
\end{abstract}




\section{Introduction}

Fluid-structure interaction (FSI) is the mutual interaction between a deformable structure and a fluid flow. FSI appears in many engineering areas, such as aeronautical, biomedical and construction engineering. Some examples of FSI include 
aero-elastic flutter of aircraft wings \citep{kamakoti2004fluid},
closure and reopening of pulmonary airways \citep{heil2011fluid},
fluid mechanics of heart valves \citep{sotiropoulos2016fluid},
flow-induced vibrations of pipes and cables \citep{nakamura2008flow},
sloshing in partially-filled containers \citep{rebouillat2010fluid}, and
self-sustained oscillations in musical instruments \citep{fabre2012aeroacoustics}.

Many of such FSI examples led to the investigation of fundamental FSI problems: simplified problems that retain important physics and help much in understanding. Examples of such classical problems are 
the flow past a freely vibrating cable \citep{newman1997direct},
the flow in collapsible tubes \citep{grotberg2004biofluid},
a flexible pipe conveying incompressible fluid \citep{xie2016flow}, and
lubrication of soft viscoelastic solids \citep{pandey2016lubrication}.
Another extensively-studied classical FSI problem is the stability of flow over compliant walls \citep{carpenter1986hydrodynamic, kumaran1995stability}, which has mainly been studied for two reasons: delay of transition to turbulence in laminar flows, and drag reduction in turbulent flows. Anisotropic, viscoelastic and permeable compliant walls have been investigated as well \citep{yeo1990hydrodynamic, hamadiche2004spatiotemporal, pluvinage2014instabilities}.

In this paper we investigate another fundamental FSI problem, namely the oscillatory laminar shear flow over a compliant viscoelastic layer on a rigid base. The compliant coating is solely driven by shear forces; normal stresses are absent. This relatively simple problem has an analytical solution. The current paper focuses on the dynamics in absence of instabilities. Specifically, it investigates how the flow dynamics changes as a result of the coupling to a viscoelastic solid.

The motivation for studying this problem comes from the possible use of compliant coatings for turbulent drag reduction, as reviewed by \cite{gad2002compliant}. Compliant coatings can delay transition to turbulence in laminar flows, thereby reducing drag. In addition, there are some indications from experiments that compliant walls can also reduce drag in turbulent flows \citep{lee1993investigation, choi1997}. However, detailed, carefully conducted and independently verified experimental studies are very scarce.
Semi-analytical studies have been performed as well. Some investigations describe dispersion relations for waves on (visco)elastic layers \citep{gad1984interaction, duncan1985dynamics, kulik2008wave, vedeneev2016propagation}. Other studies consider the response of viscoelastic layers to travelling pressure pulses or waves \citep{duncan1986response, kulik2012action}.
Finally, several numerical studies have appeared the past two decades. The resolvent formulation was used to consider the interactions between a compliant wall and turbulence \citep{luhar2015framework, luhar2016design}. In addition, some Direct Numerical Simulations (DNSs) of turbulent flow over compliant walls have been performed \citep{endo2002direct, xu2003turbulence, fukagata2008evolutionary, kim2014space}. The walls were modelled as spring-damper-supported plates or membranes. The surface motion was restricted to the vertical direction in most studies. There is still a need for DNSs of turbulent flow over a truly viscoelastic layer, which is more appropriate to model realistic coatings \citep{kulik2008wave}.

The present problem also relates to the field of physiological fluid mechanics, with hemodynamics (the dynamics of blood flow) in particular. There are at least three characteristics that distinguish blood flow from steady flow in rigid channels: pulsatility, distensibility and viscoelasticity.
Cardiovascular flow is pulsatile: there is a periodically varying flow on top of the mean flow. An overview of pulsatile flow theory is given by \citet{gundogdu1999present1}. The flow is oscillatory when there is zero mean flow. Some classical papers about oscillating flows were written by Womersley and co-workers \citep{womersley1955method, hale1955velocity}. Here we also consider an oscillatory flow.
Distensibility refers to the characteristic that an increase of the intravascular pressure results in swelling of the blood vessel, i.e. the vessel radius increases. Recent work shows that the axial (or longitudinal) displacement of arterial walls might be significant as well under certain conditions \citep{canic2014fluid}. Distensibility is not relevant in the present problem, as normal stresses are absent. The focus is on axial displacements of the viscoelastic wall.
Finally, blood vessels are viscoelastic, see e.g. \cite{bergel1961dynamic} and \cite{canic2014fluid} (with references therein). Viscoelastic materials exhibit both elastic and viscous behaviour. The inclusion of viscoelasticity is important for predicting the correct hemodynamics \citep{valdez2009analysis}. The present study therefore includes viscoelasticity.




This paper is organized as follows. Section \ref{sec:theory} describes the relevant theory related to FSI and solid viscoelasticity. Section \ref{sec:analytical_solution} introduces the specific FSI problem and derives the analytical solution. The dynamics of the coupled fluid-solid system is considered in section \ref{sec:dynamics}. In section \ref{sec:interface_quantities} we investigate how the interface velocity and shear stress change due to the viscoelastic coating. The paper closes with the conclusions and a discussion in section \ref{sec:discussion}.

\section{Theory} \label{sec:theory}
This section provides the relevant theory for the present FSI problem. The first subsection introduces the general FSI problem with the corresponding equations of motion and interface conditions. The second subsection summarizes the theory of structural viscoelasticity.

\subsection{General FSI problem}
Consider a volume that encloses both fluid and solid with the associated fluid-structure interface. The equations of motion for the fluid are:
\begin{align}
\rhof \D{u_i^f}{t} & = \pd{\stresstenf{ij}}{x_j} + \rhof f_i^f, \label{eq:fluid_momentum_equation} \\
\frac{1}{\rhof} \D{\rhof}{t} & = -\pd{u_j^f}{x_j}, \label{eq:fluid_mass_conservation}
\end{align}
with material (or total) time derivative $\mathrm{D}/\mathrm{D}t = \partial / \partial t + u_j^f \partial / \partial x_j$, velocity $u_i$, time $t$, spatial coordinate $x_j$, density $\rho$, stress tensor $\stress{ij}$ and body force $f_i$. The super- or subscripts $f$ indicate the fluid phase. The Einstein summation convention for repeated indices is used. Equation \ref{eq:fluid_momentum_equation} expresses momentum transport and equation \ref{eq:fluid_mass_conservation} denotes mass conservation. 
The fluid is assumed to be incompressible and Newtonian, which gives the following constitutive relation for the fluid stress:
\begin{align}
\stresstenf{ij} = -p \delta_{ij} + \mu \lb \pd{u_i^f}{x_j} + \pd{u_j^f}{x_i} \rb, \label{eq:fluid_stress}
\end{align}
with dynamic viscosity $\mu$ and Kronecker delta function $\delta_{ij}$. The pressure $p$ enforces the incompressibility condition given by $\partial u_j^f / \partial x_j = 0$, such that $\rhof$ is a constant.

The fluid exerts stresses on the solid at the interface, which results in structural deformations. Let $\bm{x}$ denote the Lagrangian coordinate vector, which is the original position of a solid particle in the undeformed medium. In a deformed medium, the particle's position becomes $\bm{X} = \bm{x} + \bm{\disp}$ with components $X_i(\bm{x},t) = x_i + \dispi{i}(\bm{x},t)$. The displacement or deformation vector $\bm{\disp}$ has components $\dispi{i}$. The particle's velocity $u_i^s(\bm{x},t)$ is the total time derivative of its actual position: $u_i^s = \dif X_i / \dif t = \partial \dispi{i} / \partial t$. The equations of motion for the solid are:
%
%
%
%
\begin{align}
\rhos \pd{u_i^s}{t} & =  \pd{\stresstens{ij}}{x_j} + \rhos f_i^s, \label{eq:solid_equation_of_motion} \\
\frac{\deltarho}{\rhos} & = - \pd{\dispi{j}}{x_j}, \label{eq:solid_mass_conservation}
\end{align}
with density change $\deltarho$ due to local compression or expansion of the solid \citep{lautrup2011physics}. The super- or subscripts $s$ indicate the solid phase. The first equation represents a momentum balance, analogous to equation \ref{eq:fluid_momentum_equation}. The second equation expresses mass conservation, analogous to equation \ref{eq:fluid_mass_conservation}. The present study is restricted to slowly varying displacement fields, which satisfy \citep{lautrup2011physics}:
\begin{equation}
\begin{aligned}
\abs{\pd{\dispi{i}(\bm{x},t)}{x_j}} \ll 1 \medtab \textrm{for all } i, j, \bm{x}, t.
\end{aligned}
\label{eq:solid_slowly_varying_displacement}
\end{equation}
That allows us to ignore density changes and non-linear deformations. A model for the solid stress tensor $\stresstens{ij}$ is presented in the next subsection.

Finally, the fluid and solid are coupled at the interface by the following kinematic and dynamic boundary conditions:
\begin{subequations}
\begin{align}
u_i^f & = u_i^s, \label{eq:kinematic_BC} \\
\stresstenf{ij}n_j & = \stresstens{ij}n_j, \label{eq:dynamic_BC}
\end{align}
\end{subequations}
where $n_j$ is a unit vector normal to the interface. Effects of surface tension are neglected.

\subsection{Structural viscoelasticity}

\subsubsection{Constitutive equation}
To close the structural equations, one needs a constitutive model that relates the stress tensor $\stresstens{ij}$ to Cauchy's (infinitesimal) strain tensor $\strain{ij}$ defined as:
\begin{equation}
\begin{aligned}
\strain{ij} = \frac{1}{2}\lb \pd{\dispi{i}}{x_j} + \pd{\dispi{j}}{x_i} \rb.
\end{aligned}
\label{eq:definition_strain_tensor}
\end{equation}
%
For linear time-translation-invariant homogeneous isotropic media, the constitutive stress-strain relation can be written in integral form as \citep{robertsson1994viscoelastic, carcione2015wave}:
\begin{equation}
\begin{aligned}
\stresstens{ij} & = \dot{\dilatmodrelax} \conv \strain{kk}\delta_{ij} + 2\dot{\shearmodrelax}\conv\strain{ij},
\end{aligned}
\label{eq:constitutive_stress_strain_relation}
\end{equation}
where the dot denotes a time derivative and the asterisk symbolizes convolution:
\begin{align}
f(t) \conv g(t) \equiv \intn{-\infty}{\infty}{f(\tau)g(t-\tau)}{\tau}.
\end{align}
The convolution expresses that the stress depends on the strain history (assuming causality).
The constitutive equation contains two relaxation functions, namely $\dilatmodrelax(t)$ for dilatation and $\shearmodrelax(t)$ for shear. A relaxation function describes how stress decays as a function of time in response to a unit step in strain. So, the relaxation functions are step responses, while their time derivatives are impulse responses. 

For purely elastic media, the stress responds immediately to changes in strain. When the strain is a step function (denoted by $\step{t}$), so is the stress. In that case, the relaxation functions are simply $\dilatmodrelax = \dilatmod \step{t}$ and $\shearmodrelax = \shearmod \step{t}$. Note that the dilatational modulus $\dilatmod$ and the shear modulus $\shearmod$ are the elastic Lam\'{e} constants. As $\dot{\stepsymbol}(t) = \delta(t)$, the Dirac delta function, the above stress-strain relation reduces to:
\begin{equation}
\begin{aligned}
\stresstens{ij} = \dilatmod\strain{kk}\delta_{ij} + 2\shearmod\strain{ij},
\end{aligned}
\end{equation}
which is well-known from linear elasticity.

\subsubsection{Viscoelastic wave equations}

Equations \ref{eq:solid_equation_of_motion}, \ref{eq:definition_strain_tensor} and \ref{eq:constitutive_stress_strain_relation} can be combined in a single viscoelastic wave equation:
\begin{equation}
\begin{aligned}
\rhos \pds{\dispi{i}}{t} = \lb \dot{\dilatmodrelax} + \dot{\shearmodrelax} \rb \conv \pd{}{x_i}\lb\pd{\dispi{k}}{x_k}\rb + \dot{\shearmodrelax} \conv \pds{\dispi{i}}{x_j} + \rhos f_i^s.
\end{aligned}
\label{eq:single_viscoelastic_wave_equation}
\end{equation}
One generally distinguishes two wave types, namely compressional waves (also called primary or P-waves) and shear waves (also called secondary or S-waves). P-waves are described by an equation for the dilatation $\bm{\diver \disp} = \strain{kk}$. S-waves are described by an equation for $\bm{ \curl \disp}$.

\subsubsection{Complex moduli}

The viscoelastic properties of a medium are fully specified by two relaxation functions. One might also provide the related complex moduli, which are especially useful in harmonic problems.
First, the Fourier transform pair is defined:
\begin{subequations}
\begin{align}
\fts{\psi(t)} & = \Psi(\omega) =               
\intn{-\infty}{+\infty}{\psi(t)\expp{-\comp \omega t}}{t}, \\
\ftinvs{\Psi(\omega)} & = \psi(t) = 
\frac{1}{2\pi}\intn{-\infty}{+\infty}{\Psi(\omega)\expp{\comp \omega t}}{\omega}.
\end{align}
\end{subequations}
Let $\relaxmod_M(t)$ be a relaxation function, then the complex modulus $\complexmod(\omega)$ is defined as its $\comp \omega$-multiplied Fourier transform \citep{tschoegl2002poisson, carcione2015wave}:
\begin{align}
\complexmod(\omega) \equiv \fts{\dot{\relaxmod}_M} = \comp \omega \fts{\relaxmod_M}.
\end{align}
The complex modulus can be written as a complex number with either amplitude and phase, or real and imaginary part:
\begin{subequations}
\begin{align}
\complexmod(\omega) & = \abs{\complexmod} \expp{\comp \lossangle} = 
\complexmodreal + \comp \complexmodimag, \\
\dissfac(\omega) & = \hspace{0mm} \frac{\complexmodimag}{\complexmodreal} \hspace{0mm} = 
			 \tan( \lossangle ),
\end{align}
\end{subequations}
with the modulus magnitude $\abs{\complexmod}$, loss angle $\lossangle$, storage modulus $\complexmodreal$, loss modulus $\complexmodimag$ and dissipation factor $\dissfac$. These quantities depend in general on frequency.
$\complexmodreal$ is a measure for stored strain energy, while $\complexmodimag$ is a measure for the rate of energy dissipation \citep{carcione2015wave}. The loss angle $\lossangle$ represents the phase shift between stress and strain. The dissipation factor is also called `loss factor' \citep{carfagni1998loss} or `loss tangent' (which is the tangent of the loss angle) \citep{pipkin1986lectures}.

\subsubsection{Steady state harmonic conditions}
When the boundary conditions and body forces of a viscoelastic problem are steady state harmonic functions of time, all field variables will have the same time dependence \citep{christensen1982theory}. Suppose that all variables have a harmonic dependence of the form $\expp{\comp \omeganot t}$. Exponentials behave nicely under convolution:
%
\begin{equation}
\begin{aligned}
f(t) \conv \expp{\comp \omeganot t} = F(\omeganot) \expp{\comp \omeganot t},
\end{aligned}
\end{equation}
where $F(\omega) = \fts{f(t)}$. Using this property and the definition of a complex modulus, the viscoelastic stress-strain relation (\ref{eq:constitutive_stress_strain_relation}) and the viscoelastic wave equation (\ref{eq:single_viscoelastic_wave_equation}) reduce to:
\begin{subequations}
\label{eq:viscoelastic_equations_for_harmonic_time_dependence}
\begin{align}
\stresstens{ij} & = \dilatmod(\omeganot) \strain{kk}\delta_{ij} + 2\shearmod(\omeganot)\strain{ij}, \\
\rhos \pds{\dispi{i}}{t} & = \Big( \dilatmod(\omeganot) + \shearmod(\omeganot) \Big) \pd{}{x_i}\lb\pd{\dispi{k}}{x_k}\rb + \shearmod(\omeganot) \pds{\dispi{i}}{x_j} + \rhos f_i^s,
\end{align}
\end{subequations}
where $\dilatmod$ and $\shearmod$ are the complex dilatational and shear moduli. These equations are the same as used in linear elasticity, except for the use of complex moduli.

\section{Analytical solution} \label{sec:analytical_solution}

\begin{figure}
\centering
\begin{minipage}[t]{0.65\linewidth}
\centering
\includegraphics[width=\textwidth]{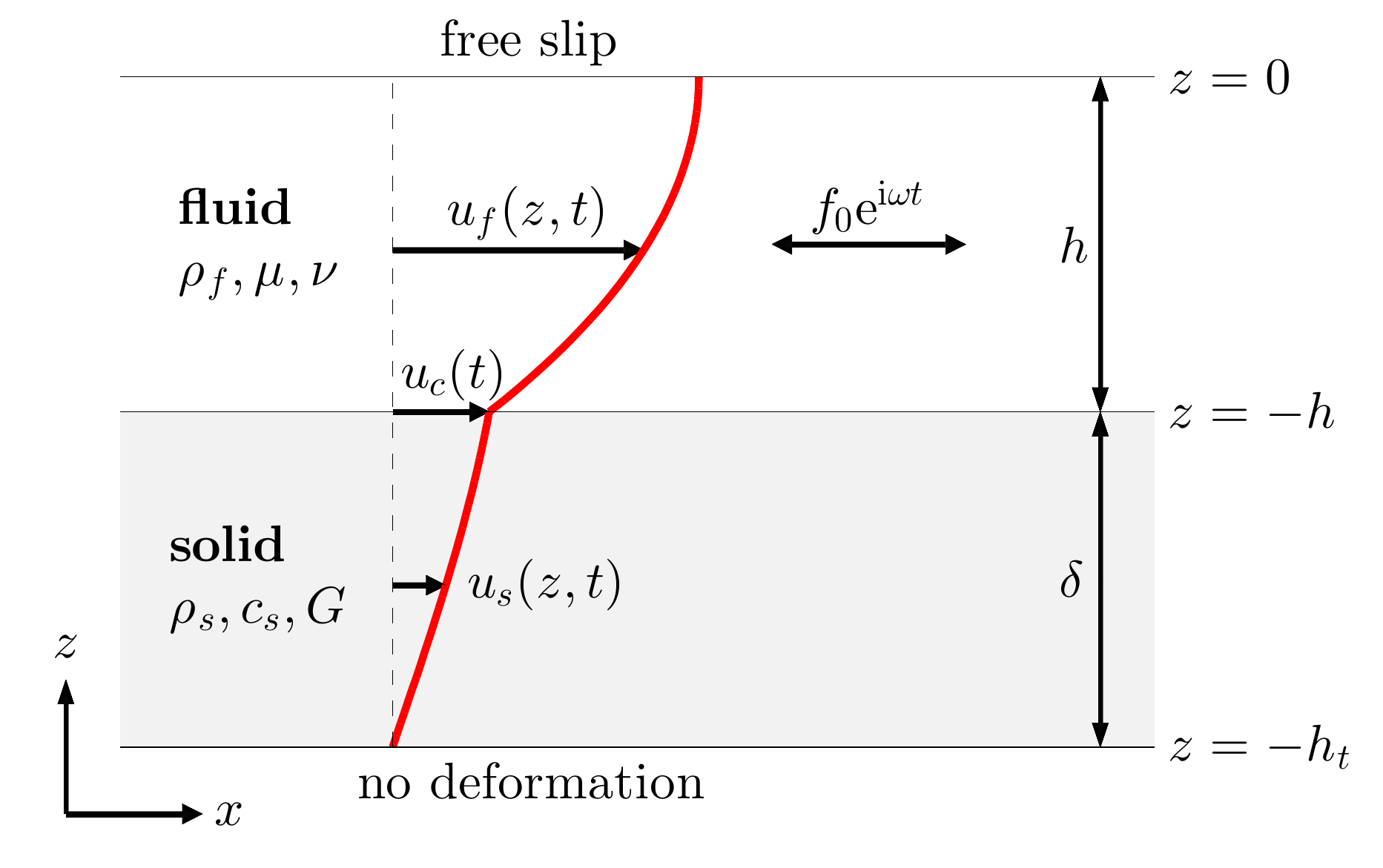}
\end{minipage}
\caption{Sketch of the fluid-structure interaction problem. A one-dimensional laminar flow is driven by a harmonic forcing. The shear stresses from the flow result in a deformation of the underlying viscoelastic coating. The fluid is indicated by a white background colour. The light-grey background colour represents the solid. The thick solid line is a velocity profile in solid and fluid.
}
\label{fig:sketch_analytical_problem}
\end{figure}

This section applies the general theory from the previous section to an oscillatory laminar flow over a viscoelastic coating. The situation is sketched in figure \ref{fig:sketch_analytical_problem}. A viscous and incompressible Newtonian fluid flows through a channel with half-height $\halfc$. The fluid has density $\rhof$, dynamic viscosity $\mu$ and kinematic viscosity $\nu = \mu/\rhof$. The flow is assumed to be laminar, one-dimensional, homogeneous in the streamwise direction, and symmetric with respect to the channel centreline ($z = 0$). The symmetry is imposed by a free-slip condition at $z = 0$. The flow is driven by a harmonic forcing with angular frequency $\omega$:
\begin{equation}
\begin{aligned}
\fxf = \fnot \expp{\comp \omega t} = \fnot \expp{\comp \phaseang}, \medtab
\phaseang = \omega t,
\end{aligned}
\end{equation}
where $\fxf$ is the flow forcing in the $x$-direction (in units $[ \unit{m \, s^{-2}} ]$) and $\phaseang$ the phase angle. The forcing amplitude $\fnot$ is a real and positive constant. Note that the complex notation is used. Physically relevant quantities are obtained by taking the real part, denoted as $\real{...}$.
Stresses from the flow result in a deformation of the underlying coating. The flow pressure is uniform and assumed to be zero ($p = 0$), so the structure is solely driven by shear forces. The coating of thickness $\tc$ is linear, time-translation-invariant, homogeneous, isotropic and viscoelastic. It is attached at the bottom to a rigid wall. It has density $\rhos$, shear-wave speed $\cs$ and shear modulus $\shearmod = \rhos \cs^2$. The total half-channel height, including the solid layer, is $\htot = \halfc + \tc$.

%


This problem allows an analytical solution, in which the following five dimensionless parameters will appear:
\begin{equation}
\begin{aligned}
\text{density ratio} \medtab 
\rhor & = \frac{\rhof}{\rhos} && \hspace{-0.3cm} 
= \frac{\text{fluid density}}{\text{solid density}}, \\
\text{geometrical parameter} \medtab 
\hr & = \frac{\halfc}{\tc} && \hspace{-0.3cm} 
= \frac{\text{fluid geometrical length-scale}}{\text{solid geometrical length-scale}}, \\
\text{frequency relative to fluid} \medtab
\omegarf & = \frac{\omega \halfc^2}{\nu} && \hspace{-0.3cm} 
= \frac{\text{forcing frequency}}{\text{frequency of viscous diffusion}}, \\
\text{frequency relative to solid} \medtab 
\omegars & = \frac{\omega \tc}{\cs} && \hspace{-0.3cm} 
= \frac{\text{forcing frequency}}{\text{frequency of shear waves}}, \\
\text{dimensionless forcing amplitude} \medtab
\fr & = \frac{\rhof \fnot \halfc}{\rhos \cs^2} && \hspace{-0.3cm} 
= \frac{\text{fluid shear stress}}{\text{resistance of solid to shear}}.
\end{aligned}
\label{eq:dimensionless_numbers}
\end{equation}
Note that there are two frequency-related dimensionless numbers: $\omegarf$ is relative to a typical fluid frequency and $\omegars$ is relative to a typical solid frequency. The parameter $\omegarf$ relates to the Womersley number $\Wo$ according to $\omegarf = \Wo^2$ \citep{womersley1955method}. The parameter $\omegars$ is a reduced or dimensionless wave number, because it equals $\omegars = \ks \tc$ with wave number $\ks = \omega/\cs$ for shear waves.


For a viscoelastic coating, the shear modulus becomes complex and frequency dependent (see previous section), i.e. $\shearmod = \shearmod(\omega)$. As this paper considers only one forcing frequency, there is no need to describe the frequency dependence of the complex modulus. At the prescribed forcing frequency, the shear modulus is simply a complex number with modulus $\abs{\shearmod}$ and phase $\phase{\shearmod}$. Derived quantities, like $\cs$ and $\omegars$, can be written in a similar way:
\begin{equation}
\begin{aligned}
\shearmod & = \abs{\shearmod} \expp{\comp \phase{G}}, \\
\cs & = \abs{\cs} \expp{\comp \phase{\cs}}, && \abs{\cs} = \sqrt{ \frac{\abs{\shearmod}}{\rhos}}, && \phase{\cs} = \frac{\phase{G}}{2}, \\
\omegars & = \abs{\omegars} \expp{\comp \phase{\omegars}}, && \abs{\omegars} = \frac{\omega \tc}{\abs{\cs}}, && \phase{\omegars} = -\phase{\cs} = -\frac{\phase{G}}{2}.
\end{aligned}
\end{equation}
The angle $\phase{\shearmod}$ is the loss angle. It has been found empirically that it always lies between 0 and $90\degree$ \citep{pipkin1986lectures}. As a result, the loss tangent $0 \leq \tan(\phase{\shearmod}) < \infty$. However, a loss tangent equal to one is considered outstandingly high \citep{chung2001review}. Therefore, the range $0 \leq \phase{\shearmod} \leq 45\degree$ might be more realistic.




The sketched problem is first solved analytically for the fluid flow. Given the aforementioned assumptions, the fluid stress (equation \ref{eq:fluid_stress}) and the momentum equations (\ref{eq:fluid_momentum_equation}) become:
%
\begin{align}
\stresstenf{13} = \mu \pd{\uf}{z}, \medtab
\pd{\uf}{t} = \nu \pds{\uf}{z} + \fxf.
\label{eq:unsteady_Stokes}
\end{align}
%
The latter equation is the unsteady Stokes equation. Because both fluid and solid behave linearly, all quantities have the same harmonic time dependence as the flow forcing. For example, the fluid velocity $\uf$, the velocity at the fluid-structure interface $\usurf$, and the solid velocity $\us$ satisfy: 
\begin{equation}
\begin{aligned}
\uf(z,t) 		& = \ufnot(z) \expp{\comp \omega t}, \\
\usurf(t) 		& = \usurfnot \expp{\comp \omega t}, \\
\us(z,t) 		& = \usnot(z) \expp{\comp \omega t}.
\end{aligned}
\end{equation}
Next, quantities are nondimensionalized as follows:
\begin{equation}
\begin{aligned}
\zbreve = \frac{z}{\halfc}, \medtab
\ubreve = \frac{u}{\uft}, \medtab 
\uft = \frac{\fnot}{\omega}.
\end{aligned}
\end{equation}
%
The velocity is normalized with a typical forcing velocity $\uft$.
The equation to be solved and its solution are:
\begin{equation}
\begin{aligned}
\pds{\ufnotbreve}{\zbreve} - \comp \omegarf \ufnotbreve + \omegarf = 0, \medtab
\left. \ufnotbreve \right|_{\zbreve = -1} = \usurfnotbreve, \medtab
\left. \pd{\ufnotbreve}{\zbreve} \right|_{\zbreve = 0} = 0,
\end{aligned}
\end{equation}
%
%
\begin{equation}
\begin{aligned}
\ufnotbreve = -\comp 
\left\lbrace 1 - \lb  1 + \frac{\usurfnotbreve}{\comp} \rb 
\frac{\cosh{\lb \sqrt{\comp \omegarf} \, \zbreve \, \rb}}{\cosh{\lb \sqrt{\comp \omegarf} \, \rb}} \right\rbrace, \medtab -1 \leq \zbreve \leq 0.
\end{aligned}
\end{equation}
%
%
%

As a next step, the deformation of the structure is derived. The present paper considers a one-dimensional problem with $\bm{\disp} = \disp(z,t) \bm{\hat{x}}$, where $\bm{\hat{x}}$ denotes a unit vector in the $x$-direction. Hence, $\bm{\diver \disp} = \strain{kk} = 0$ and only shear waves will appear. The only non-zero strains and stresses are $\strain{13} = \strain{31}$ and $\stress{13} = \stress{31}$. Given the harmonic time dependence and the absence of body forces ($f_i^s = 0$), equations \ref{eq:viscoelastic_equations_for_harmonic_time_dependence} reduce to:
%
\begin{equation}
\begin{aligned}
\stresstens{13} = \shearmod \pd{\disp}{z}, \medtab
\pds{\disp}{t} = \frac{\shearmod}{\rhos} \pds{\disp}{z}. \\
\end{aligned}
\end{equation}
The latter equation is a wave equation with wave speed $\cs = \sqrt{G/\rhos}$.
A time derivative of that equation yields the following equation for the structural deformation velocity $\us(z,t)$:
\begin{equation}
\begin{aligned}
\pds{\us}{(z/\tc)} + \omegars^2 \us = 0, \medtab
\left. \us \right|_{z = -\htot} = 0, \medtab
\left. \us \right|_{z = -\halfc} = \usurf.
\end{aligned}
\end{equation}
This equation can be solved easily for $\us$. The displacement then follows from the relation $\us = \partial \disp / \partial t = \comp \omega \disp$:
\begin{equation}
\begin{aligned}
\us = \usurf \frac{\sin{\lb \omegars \lb \frac{z+\htot}{\tc} \rb \rb}}{\sin{\lb \omegars \rb}}, 
\medtab \disp = \frac{\us}{\comp \omega},
\medtab -\htot \leq z \leq -\halfc.
\end{aligned}
\end{equation}

Both the fluid and solid solution now satisfy the kinematic boundary condition, namely the continuity of velocity (equation \ref{eq:kinematic_BC}). However, they should also satisfy the dynamic boundary condition, namely continuity of stress (equation \ref{eq:dynamic_BC}). Specifically, at the fluid-structure interface $\stresstenf{13} = \stresstens{13}$ or:
\begin{equation}
\begin{aligned}
\tauw = \mu \pd{\uf}{z} = \shearmod \pd{\disp}{z} \mathrm{\medtab or \medtab}
- \frac{\rhor \, \hr \, \omegars^2}{\comp \omegarf} \pd{\ufnotbreve}{(z/\halfc)} = \pd{\usnotbreve}{(z/\tc)},
\medtab \text{at } z = -\halfc.
\end{aligned}
\label{eq:stress_interface_relation}
\end{equation}
That gives an equation which can be solved for $\usurfnot$, yielding:
%
%
%
\begin{equation}
\begin{aligned}
\frac{\usurfnot}{\uft} = \frac{- \comp \, \intpar}{\intpar - 1}, \medtab
\intpar 		= \rhor \, \hr \, \omegars \tan{\lb\omegars\rb} \frac{ \tanh{\lb \sqrt{\comp \omegarf} \rb} }{ \sqrt{\comp \omegarf} },
\end{aligned}
\end{equation}
%
where $\intpar$ is a new dimensionless parameter, which we call the shear interaction parameter. Note that in general $\intpar$ is a complex number. The fluid and solid velocity can then be rewritten as:
%
%
\begin{subequations}
\begin{align}
\frac{\uf(z,t)}{\uft} 	& = - \comp \,
\left\lbrace 1 - \lb  \frac{1}{1 - \intpar} \rb 
\frac{\cosh{\lb \sqrt{\comp \omegarf} \, z/\halfc \, \rb}}{\cosh{\lb \sqrt{\comp \omegarf} \, \rb}} \right\rbrace \expp{\comp \omega t}, \medtab && -h \leq z \leq 0, \label{eq:fluid_velocity_solution} \\
\frac{\us(z,t)}{\uft} & = \frac{- \comp \, \intpar}{\intpar - 1} \frac{\sin{\lb \omegars \lb \frac{z+\htot}{\tc} \rb \rb}}{\sin{\lb \omegars \rb}} \expp{\comp \omega t}, 
\medtab && -\htot \leq z \leq -\halfc.
\end{align}
\end{subequations}
%
Although the whole problem is governed by five dimensionless parameters, the interaction of fluid and solid is governed by one parameter $\intpar$. In turn, this parameter depends on all dimensionless numbers in equation \ref{eq:dimensionless_numbers}, except for $\fr$. The dimensionless forcing amplitude $\fr$ only appears in $\uft$. Note that both fluid and solid velocity are linearly proportional to $\uft$ and $\fr$, such that a stronger forcing also yields larger velocities in fluid and solid.

For the interpretation of the obtained solution, it will be convenient to consider $\intpar$ in the limits of small and large $\omegarf$. When $\omegarf$ approaches zero, $\tanh \lb \sqrt{\comp \omegarf} \rb /\sqrt{\comp \omegarf} \approx 1$. When $\omegarf$ goes to infinity, $\tanh \lb \sqrt{\comp \omegarf} \rb \approx 1$. Hence, the following expressions for $\intpar$ are obtained:
\begin{subequations}
\begin{alignat}{2}
\omegarf & \goto 0 \medtab &&
\intpar = \rhor \, \hr \, \omegars \tan{\lb\omegars\rb}, \label{eq:intpar_small_omegarf} \\
\omegarf & \goto \infty \medtab &&
\intpar = \frac{ \rhor \, \hr \, \omegars \tan{\lb\omegars\rb} }{ \sqrt{\comp \omegarf} } = 
\frac{ \rhor \, \omegars \tan{\lb\omegars\rb} }{ \sqrt{\comp \lb \frac{\omega \tc^2}{\nu} \rb} }. \label{eq:intpar_large_omegarf}
\end{alignat}
\end{subequations}
These relations will be useful in the next sections.

\begin{figure}
\centering
\begin{minipage}[t]{1\linewidth}
\centering
\includegraphics[width=\textwidth]{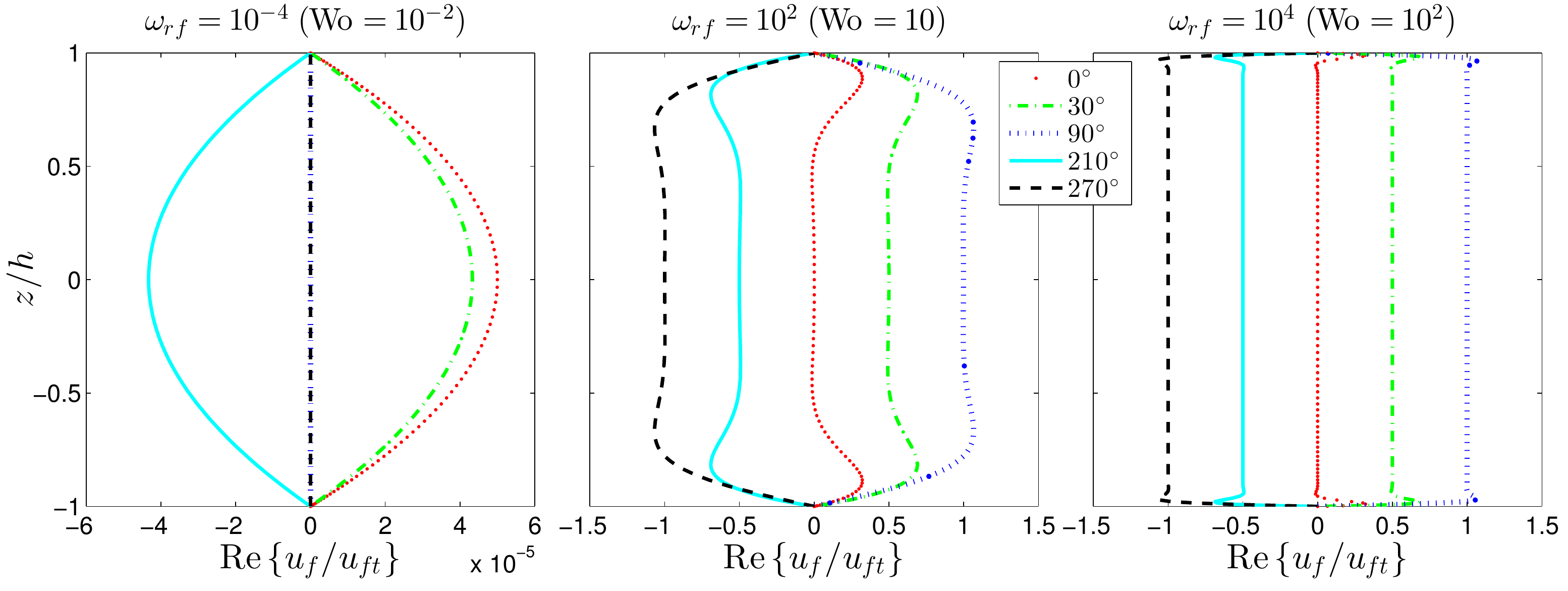}
\end{minipage}
\caption{Fluid velocity profiles for three different values of $\omegarf$ at five different phase angles $\phi = \omega t$. The interface velocity $\usurf = 0$ for all the cases. This is the only figure where the fluid velocity is also shown in the upper channel half.}
\label{fig:velocity_profiles_different_omegarf_rigid_solid}
\end{figure}

\section{Dynamics} \label{sec:dynamics}

In this section we analyse the analytical solution that has been derived above. 
The first subsection considers the dynamics in the flow, with specific attention for the parameter $\omegarf$.
The second subsection describes the dynamics in the solid. It considers the parameters $\omegars$ and $\phiG$ in particular.
The third subsection investigates the dynamics of the combined fluid-solid system in certain limiting cases.


\subsection{Dynamics in fluid} \label{sec:dynamics_fluid_only}

The flow dynamics in the absence of a compliant structure is described by $\omegarf = \omega \halfc^2 / \nu$, the related Womersley number $\Wo = \sqrt{ \omega \halfc^2 / \nu }$ or the Stokes number $\sqrt{ \omega \halfc^2 / 2 \nu }$. The latter defines the characteristic length scale $\lstokes = \sqrt{ 2\nu/\omega }$, the Stokes layer thickness. Loosely speaking, $\lstokes$ defines the extent of a near-wall region where viscous effects prevail over inertial effects.
A second important dimensionless number is the Reynolds number, which quantifies when flow instabilities might arise. For instance, turbulent bursts in oscillatory pipe flow first occur  at $\re_c \approx 800 \sqrt{\omega R^2 / \nu}$ for $2.3 \lesssim \sqrt{\omega R^2 / \nu} \lesssim 41$. Here $R$ is the pipe radius and $\re_c = u_m 2 R / \nu$ is a critical Reynolds number based on $u_m$, the amplitude of the oscillatory component of the cross-sectional mean velocity \citep{gundogdu1999present1}. The present paper considers laminar flow, so the Reynolds number should be sufficiently small.



The laminar-flow patterns can be classified into three types: quasi-steady ($\Wo \lesssim 1.3$), intermediate ($1.3 \lesssim \Wo \lesssim 28$) and inertia-dominated ($\Wo \gtrsim 28$) \citep{gundogdu1999present1}. Figure \ref{fig:velocity_profiles_different_omegarf_rigid_solid} shows the fluid velocity profiles for three different Womersley numbers, corresponding to the three different flow types. 
At very low $\Wo$, the flow is quasi-steady and dominated by viscous diffusion, so the Stokes layer thickness is much larger than the half-channel height. The unsteady Stokes equation (\ref{eq:unsteady_Stokes}) reduces to $\nu \, \partial^2 \uf / \partial z^2 + \fxf = 0$. As a result, the velocity is parabolic and perfectly in phase with the flow forcing. Furthermore, the relevant velocity scale is $\fnot h^2 / \nu = \uft \, \omegarf$. This explains why $\uf/\uft = O(\omegarf) = O(10^{-4})$ in the left subfigure.
The middle subfigure shows the velocity profiles at an intermediate Womersley number. Now $\lstokes < h$ and the viscous effects are mainly confined to a layer near the wall. Close to the channel centreline, the flow velocity satisfies $\partial u_f / \partial t = \fxf$. As a result, the velocity is approximately uniform and 90$\degree$ out of phase with the forcing. In addition, $\uft = \fnot/\omega$ is the characteristic velocity scale.
The right subfigure belongs to a large Womersley number. Viscous effects are confined to an even thinner layer close to the wall. The largest part of the flow is inertia-dominated, as is apparent from the flat velocity profile and the 90$\degree$ phase delay.

\begin{figure}
\centering
\begin{minipage}[t]{1\linewidth}
\centering
\includegraphics[width=\textwidth]{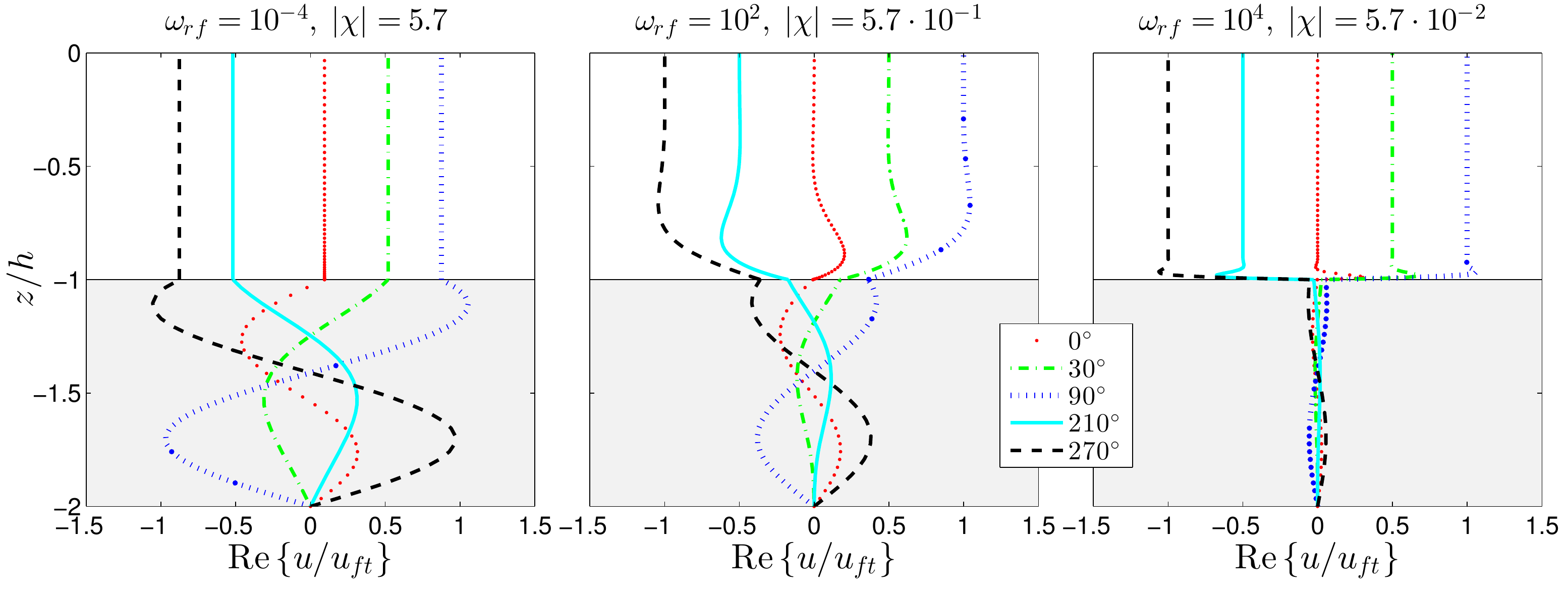}
\end{minipage}
\caption{Velocity profiles for three different values of $\omegarf$ at five different phase angles. The other parameters are $\rhor = 1$, $\hr = 1$, $\left| \omegars \right| = 1.75\pi$, $\phiG = 10\degree$.
}
\label{fig:velocity_profiles_different_omegarf}
\end{figure}

While figure \ref{fig:velocity_profiles_different_omegarf_rigid_solid} depicts the flow over a rigid wall, figure \ref{fig:velocity_profiles_different_omegarf} shows an example of the flow over a deformable wall. The velocity profiles in solid and fluid are shown for five different phase angles. The three subfigures correspond to different values of $\omegarf$ (as in figure \ref{fig:velocity_profiles_different_omegarf_rigid_solid}). Due to the deformation of the solid, the fluid has an apparent slip. For small $\omegarf$, viscous effects are strong, which results in an almost uniform velocity profile in the fluid. In addition, the flow is not any more in phase with the forcing because of the coupling to the solid. For large $\omegarf$, the interface velocity $\usurf$ is close to zero: the relatively small viscous forces can only produce small solid deformations. As a result, the fluid velocity profiles for high $\omegarf$ are very similar to the ones for flow over a rigid wall (see figure \ref{fig:velocity_profiles_different_omegarf_rigid_solid}). 


\subsection{Dynamics in solid} \label{sec:dynamics_solid_only}

The solid dynamics is governed by travelling shear waves. This can be observed more easily when the solution for $\us$ is rewritten in terms of complex exponentials:
\begin{equation}
\begin{aligned}
\us = \frac{\usurfnot}{2\comp\sin{\omegars}}
\left\lbrace   
\underbrace{\expp{ \comp \frac{\omega}{\cs} \lb z + \htot + \cs t \rb }					}_{\text{downward travelling wave}} - 
\underbrace{\expp{ -\comp \frac{\omega}{\cs} \lb z + \htot - \cs t \rb }					}_{\text{upward travelling wave}}
\right\rbrace.
\end{aligned}
\label{eq:superposition_travelling_waves}
\end{equation}
The deformation velocity thus results from the superposition of a downward and an upward travelling wave. The shear stress on the coating surface generates a downward travelling shear wave which reflects at the rigid wall and turns into an upward travelling wave. The interference of both waves results in a standing wave pattern.

\begin{figure}
\centering
\begin{minipage}[t]{1\linewidth}
\centering
\includegraphics[width=\textwidth]{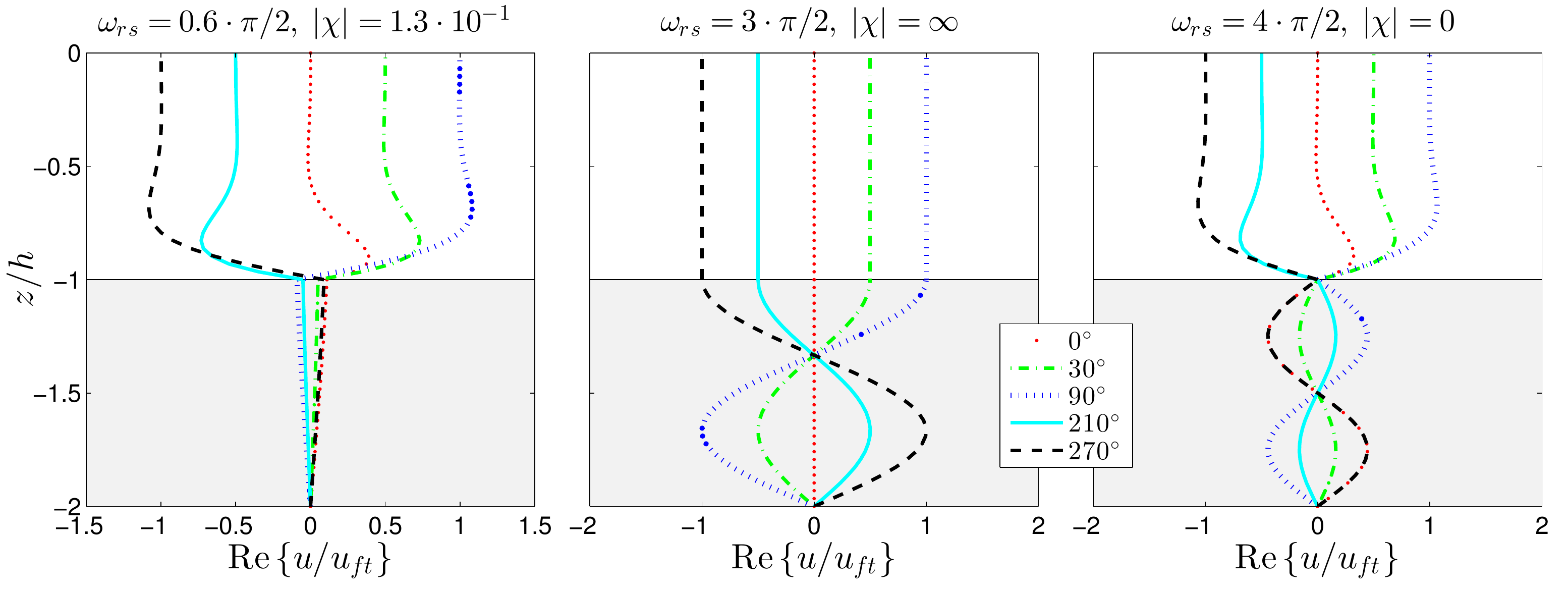}
\end{minipage}
\caption{Velocity profiles for three different values of $\omegars$ at five different phase angles. The other parameters are $\rhor = 1$, $\hr = 1$, $\omegarf = 10^{2}$, $\phiG = 0\degree$.
}
\label{fig:velocity_profiles_different_omegars}
\end{figure}

Like with strings and pipes in acoustics, an elastic solid displays resonances at certain wavelengths or frequencies. The relevant parameter is $\omegars$ or $\lambda/\tc = 2\pi/\omegars$, where $\lambda$ is the wavelength. \cite{kulik2008wave} also recognized the importance of the parameter $\omegars$, which they denoted as $\omega H / C_t$ with coating thickness $H$ and shear-wave speed $C_t$. Both odd and even modes can be distinguished. They have the following characteristics:
\begin{equation}
\begin{aligned}
\text{Odd modes}: &&
\cos{\omegars} = 0, &&
\omegars & = \frac{\pi}{2}, \frac{3\pi}{2}, \frac{5\pi}{2}, ... &&
\usurfnot = -\comp \, \uft, &
\left.\pd{\us}{z} \right|_{z = -\halfc} = 0. \\
\text{Even modes}: &&
\sin{\omegars} = 0, &&
\omegars & = \pi, 2\pi, 3\pi, ... &&
\usurfnot = 0, &
\left. \us \right|_{z = -\halfc} = 0.
\end{aligned}
\label{eq:solid_resonance_modes}
\end{equation}
Figure \ref{fig:velocity_profiles_different_omegars} shows the velocity profiles for an elastic solid and three different values of $\omegars$. 
The left subfigure corresponds to the quasi-steady limit: the solid velocity profiles depend linearly on $z$ and the deformation is small.
The centre subfigure shows an odd resonance mode. The right subfigure shows an even resonance mode.

\begin{figure}
\centering
\begin{minipage}[t]{1\linewidth}
\centering
\includegraphics[width=\textwidth]{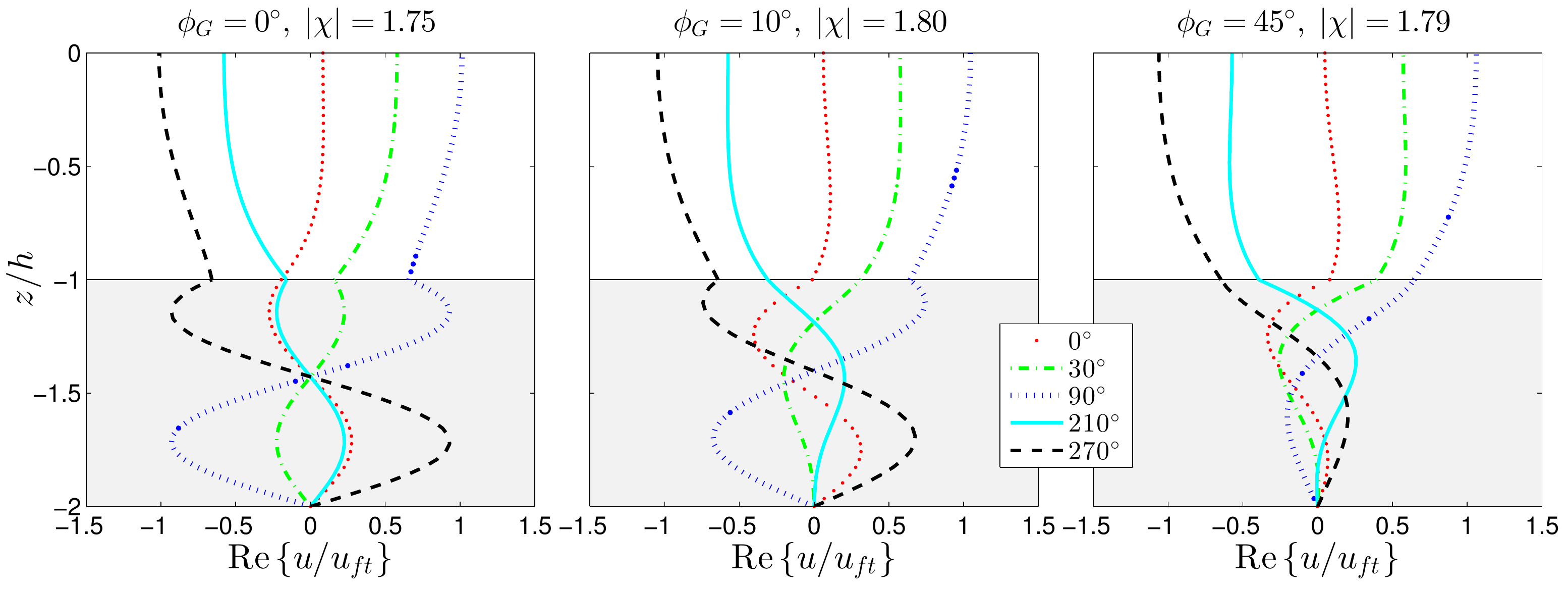}
\end{minipage}
\caption{Velocity profiles for three different damping parameters $\phase{\shearmod}$ at five different phase angles. The other parameters are $\rhor = 1$, $\hr = 1$, $\omegarf = 10$, $\left| \omegars \right| = 1.75\pi$.}
\label{fig:velocity_profiles_different_damping}
\end{figure}

While figure \ref{fig:velocity_profiles_different_omegars} displays the dynamics for a fully elastic solid, figure \ref{fig:velocity_profiles_different_damping} considers the effect of viscoelasticity. It depicts the velocity profiles for three different values of $\phiG$. Whereas the deformation is sinusoidal for $\phiG = 0\degree$, it shows exponential decay in space for non-zero $\phiG$. For increasing $\phiG$, there is more damping and the wave amplitude decreases over a shorter typical distance, in agreement with \cite{kulik2008wave}.
%
%
To quantify these findings, equation \ref{eq:superposition_travelling_waves} is rewritten to replace $\cs$, which is complex for viscoelastic media. First, the complex wave number $\ks = \omega/\cs$ is decomposed as follows:
\begin{equation}
\begin{aligned}
\ks = \frac{\omega}{\cs} = \ksp - \comp \ksa, \medtab
\ksp = \abs{\ks} \cos{\phase{\cs}}, \medtab
\ksa = \abs{\ks} \sin{\phase{\cs}}, \medtab
\phase{\cs} = \frac{\phase{G}}{2},
\end{aligned}
\end{equation}
with a real part $\ksp$ for propagation and an imaginary part $\ksa$ for attenuation. One might call $\ksp$ the (real) wave number and $\ksa$ the attenuation factor \citep{carcione2015wave}. Using these relations, equation \ref{eq:superposition_travelling_waves} can be written as:
\begin{equation}
\begin{aligned}
\us & = \frac{\usurfnot}{2\comp\sin{\omegars}}
\left\lbrace   
\underbrace{\expp{\ksa \lb z + \htot \rb} \expp{ \comp \ksp \lb z + \htot + \csp t \rb }					}_{\text{downward travelling wave}} - 
\underbrace{\expp{-\ksa \lb z + \htot \rb} \expp{ -\comp \ksp \lb z + \htot - \csp t \rb }					}_{\text{upward travelling wave}}
\right\rbrace,
\end{aligned}
\label{eq:superposition_decaying_travelling_waves}
\end{equation}
with $\csp = \omega/\ksp$ the (real) propagation velocity or phase velocity. The shear deformation of a viscoelastic medium is indeed a superposition of exponentially-decaying travelling shear-waves. One can define a characteristic decay length $\lsa = 1/\ksa$. For increasing $\phiG$, $\ksa$ increases and $\lsa$ decreases, thus confirming the above findings.

\subsection{Dynamics dominated by fluid or solid} \label{sec:dynamics_limiting_cases}

\begin{figure}
\centering
\begin{minipage}[t]{1\linewidth}
\centering
\includegraphics[width=\textwidth]{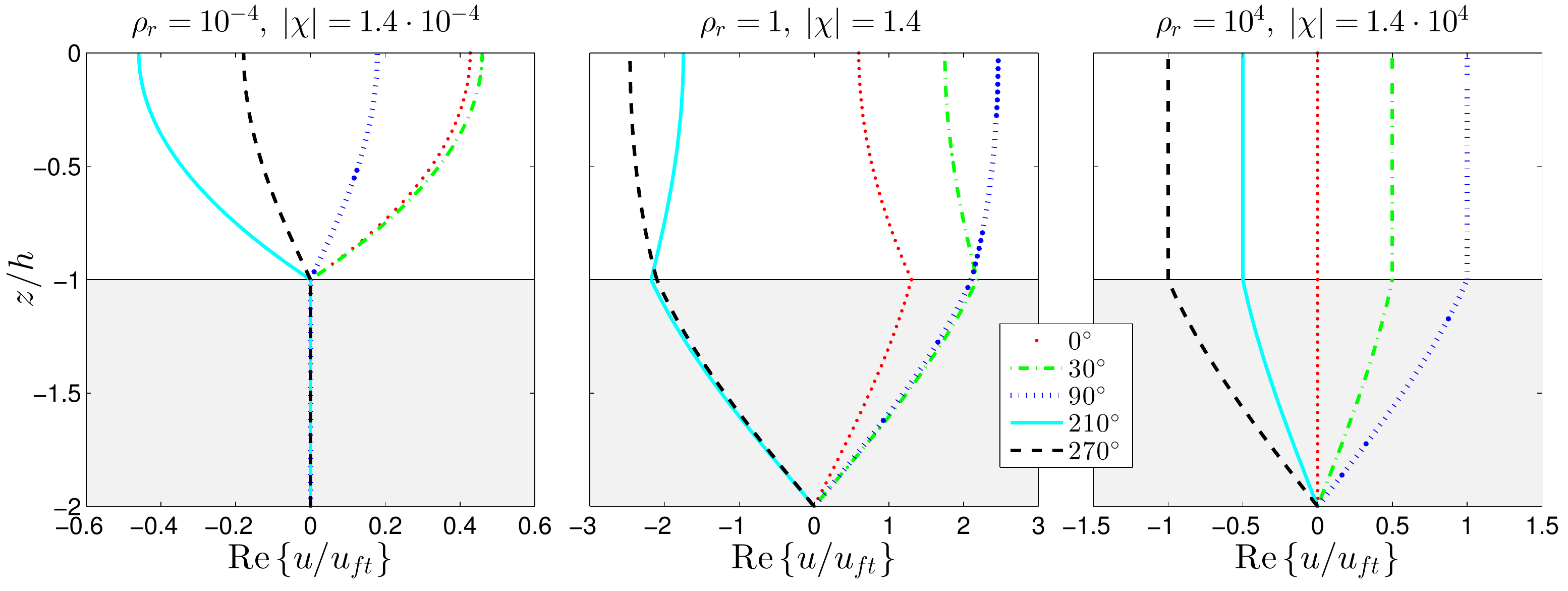}
\end{minipage}
\caption{Velocity profiles for three different density ratios $\rhor$ at five different phase angles. The other parameters are $\hr = 1$, $\omegarf = 1$, $\omegars = 1$, $\phiG = 0\degree$.
}
\label{fig:three_types_global_dynamics_density_ratio}
\end{figure}

The dynamics of the whole fluid-solid system is a complex interplay between fluid and solid dynamics. However, the interface dynamics is relatively simple in certain limiting cases. Consider for instance figure \ref{fig:three_types_global_dynamics_density_ratio}, which shows the velocity profiles for three different density ratios $\rhor$. For small $\rhor$, the solid is heavy and the interface velocity becomes zero, which corresponds to a no-slip condition for the fluid. For large $\rhor$, the solid is lightweight and the fluid velocity becomes uniform. In other words, the fluid satisfies a free-slip boundary condition at the wall. Inspired by this example, the present subsection shows how the solid can be adjusted in such a way that the fluid satisfies either a no-slip or a free-slip boundary condition.

\vspace{\baselineskip}
\textbf{No-slip condition for fluid}
The fluid behaves as shown in figure \ref{fig:velocity_profiles_different_omegarf_rigid_solid} when it satisfies a no-slip boundary condition at the wall, i.e. when the interface velocity becomes zero ($\usurfnot = 0$). This will happen in the limit $\abs{\intpar} \goto 0$. That limit can be achieved by adjusting the solid in the following ways:
%
%
%
%
\begin{itemize}[leftmargin=*,labelindent=5mm,labelsep=1mm]
	\item Use a heavy solid: $\rhos \goto \infty$ and $\rhor \goto 0$.
	\item Use a stiff solid: $\abs{\cs} \goto \infty$ and $\abs{\omegars} \goto 0$.
	\item Use a thin solid: $\tc \goto 0$, $\abs{\omegars} \goto 0$ and $\abs{\tan\omegars} \goto 0$ (while $\hr \, \omegars$ stays constant).
	\item Use an even resonance mode: adjust coating thickness and shear velocity such that $\sin\omegars = 0$ and $\tan\omegars = 0$. This is only possible for a purely elastic solid.
\end{itemize}
%
%
The interface dynamics is governed by the solid, because the solid sets the interface velocity to zero, independent of the instantaneous fluid velocity.
The no-slip condition can also be obtained for a low viscosity fluid, because $\abs{\intpar} \goto 0$ for $\omegarf \goto \infty$ (see figure \ref{fig:velocity_profiles_different_omegarf}).



\vspace{\baselineskip}
\textbf{Free-slip condition for fluid}
In the limit $\abs{\intpar} \goto \infty$, the fluid satisfies a free-slip boundary condition at the interface: $\partial \uf / \partial z = 0$ at $z = -\halfc$. The fluid velocity becomes uniform: $\uf = -\comp \uft \exp{\comp \omega t}$ (see equation \ref{eq:fluid_velocity_solution}). Consider again the situation that only the solid parameters are adjustable. The limit $\abs{\intpar} \goto \infty$ can be achieved in the following ways:
%
%
\begin{itemize}[leftmargin=*,labelindent=5mm,labelsep=1mm]
	\item Use a lightweight solid: $\rhos \goto 0$ and $\rhor \goto \infty$.
	\item Use a soft solid: $\abs{\cs} \goto 0$ and $\abs{\omegars} \goto \infty$.
	\item Use an odd resonance mode: adjust coating thickness and shear velocity such that $\cos\omegars \goto 0$ and $\tan\omegars \goto \infty$. This is only possible for a purely elastic solid.
\end{itemize}
%
%
The interface dynamics is governed by the fluid in the first two cases, because the fluid sets the interface velocity as if the solid is absent. Indeed, the same solution for $\uf$ is obtained when boundaries are absent. Equation \ref{eq:unsteady_Stokes} then reduces to $\partial \uf / \partial t = \fxf$, which is solved by $\uf = -\comp \uft \exp{\comp \omega t}$. As a result, $\uf$ lags behind $\fxf$ by 90$\degree$. This is indeed observed in figure \ref{fig:three_types_global_dynamics_density_ratio} (right), which shows an example of the free-slip boundary condition.

Note that the free-slip condition is not necessarily obtained for a thick solid. When $\tc \goto \infty$, also $\abs{\omegars} \goto \infty$, but $\hr \, \omegars$ stays constant. Therefore, $\intpar$ becomes proportional to $\tan{\lb\omegars\rb}$. It follows that $\intpar$ is periodic for real $\omegars$ and approaches a constant complex number when $\abs{\omegars} \goto \infty$ for complex $\omegars$. This will be shown in the next section.


\section{Interface velocity and shear stress} \label{sec:interface_quantities}

The previous section has provided a qualitative description of the dynamics. In addition, the behaviour in some limiting cases has been considered. However, in general the fluid satisfies neither a no-slip nor a free-slip boundary condition (as in figure \ref{fig:three_types_global_dynamics_density_ratio} centre). This section therefore quantifies the more general dynamics.
In the context of blood flow and drag reduction, the velocity and shear stress at the fluid-solid interface are interesting quantities which are considered below in two subsections. The third subsection discusses the resonances that appear in graphs of velocity and shear stress. The final subsection provides a set of parameters that could represent realistic experiments.


\begin{figure}
\centering
\begin{minipage}[t]{1\linewidth}
\centering
\includegraphics[width=\textwidth]{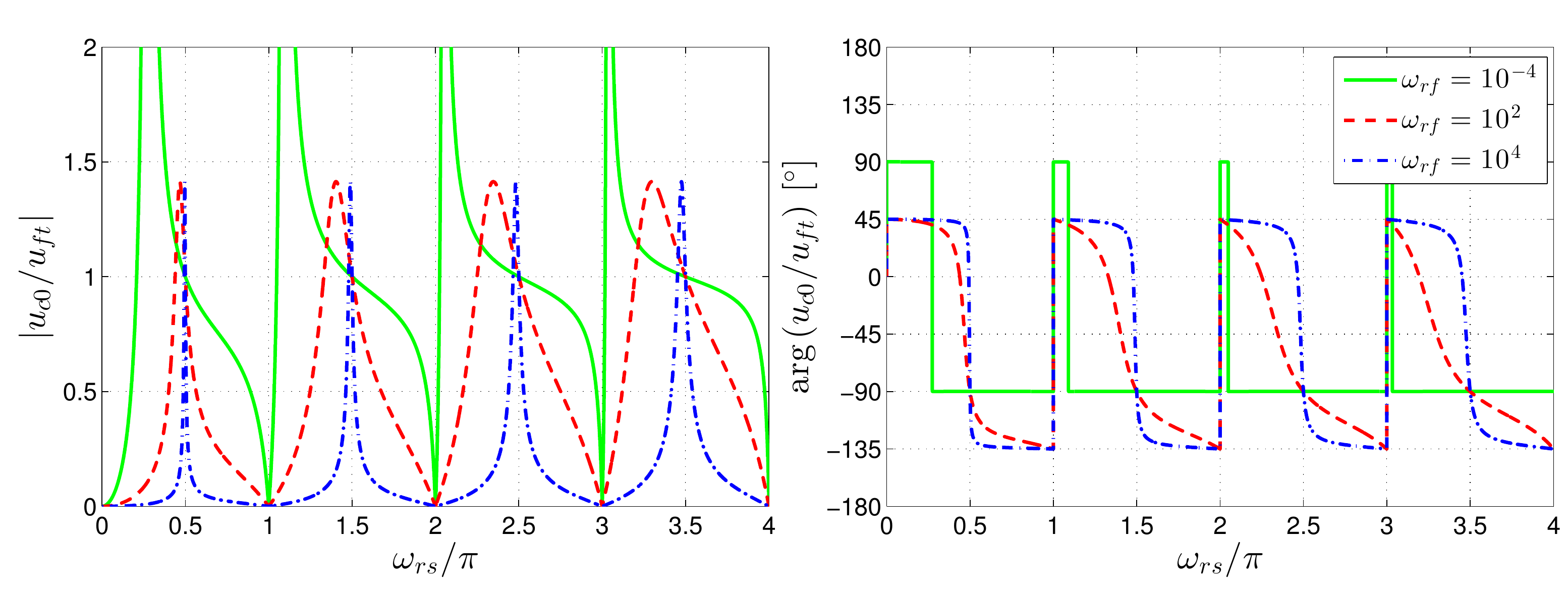}
\end{minipage}
\caption{Magnitude (left) and phase (right) of the normalized interface velocity as function of $\omegars$ for three different values of $\omegarf$ and $\rhor = 1$, $\hr = 1$, $\phiG = 0\degree$.
The increase of $\omegars$ results from a decrease of the coating stiffness.
}
\label{fig:interface_velocity_amplitude_and_phase}
\end{figure}

\begin{figure}[b!]
\centering
\begin{minipage}[t]{1\linewidth}
\centering
\includegraphics[width=\textwidth]{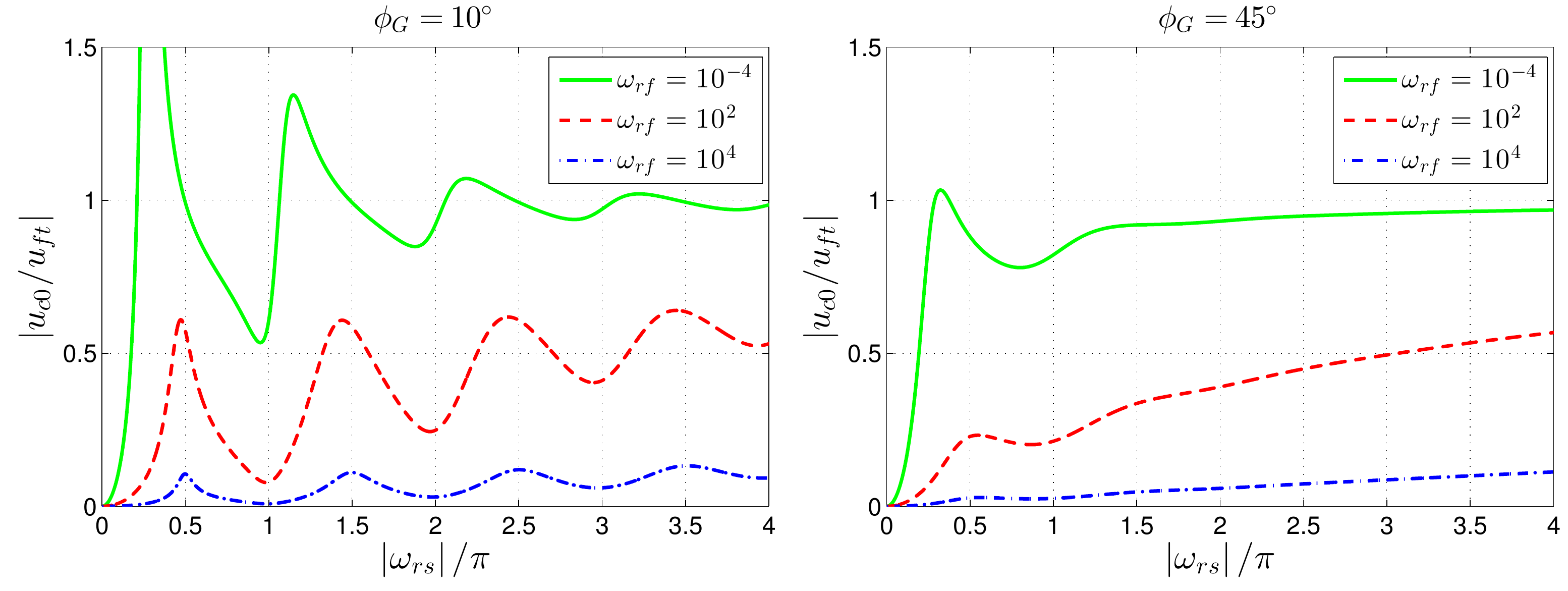}
\end{minipage}
\caption{Magnitude of the normalized interface velocity as function of $\left| \omegars \right|$ for three different values of $\omegarf$ and $\rhor = 1$, $\hr = 1$. 
The increase of $\left| \omegars \right|$ results from a decrease of the coating stiffness.
Left: low damping with $\phiG = 10\degree$. Right: high damping with $\phiG = 45\degree$.
}
\label{fig:interface_velocity_different_omegarf}
\end{figure}

\subsection{Interface velocity}

This section investigates the dependence of the normalized interface velocity $\usurfnot/\uft = -\comp \intpar / (\intpar - 1)$ on the various problem parameters.
Figure \ref{fig:interface_velocity_amplitude_and_phase} shows the amplitude and phase of the normalized interface velocity for an \textit{elastic} solid. The dependence on $\omegars$ for given $\rhor$, $\hr$ and $\omegarf$ clarifies how $\usurfnot$ changes with coating stiffness. The increase of $\omegars$ on the horizontal axis results from a decreasing $\cs$.
The solid resonance modes described in equation \ref{eq:solid_resonance_modes} are evident. The odd and even modes appear for all $\omegarf$. Specifically, the even modes have no surface velocity, while odd modes ($\usurfnot/\uft = -\comp$) have magnitude 1 and phase $-90 \degree$.

The figure also exhibits resonances that are not present in the solid system only, but result from the coupling with the fluid. These modes appear when $\real{\intpar} \approx 1$. They are especially strong for low $\omegarf$, because $\intpar$ becomes then a real number (equation \ref{eq:intpar_small_omegarf}). For $\rhor = \hr = 1$, the resonances occur at $\omegars$-values slightly larger than $\omegars = \pi, 2\pi, ...$, so close to the even solid modes.
For high $\omegarf$ (equation \ref{eq:intpar_large_omegarf}), the resonances occur when $\intpar \approx 1 - \comp$, such that $\usurfnot/\uft \approx \intpar$ with magnitude $\sqrt{2}$ and phase $-45\degree$. For $\rhor = \hr = 1$, the corresponding $\omegars$-values are slightly smaller than $\omegars = \pi/2, 3\pi/2, ...$, so close to the odd solid modes.


\begin{figure}
\centering
\begin{minipage}[t]{1\linewidth}
\centering
\includegraphics[width=\textwidth]{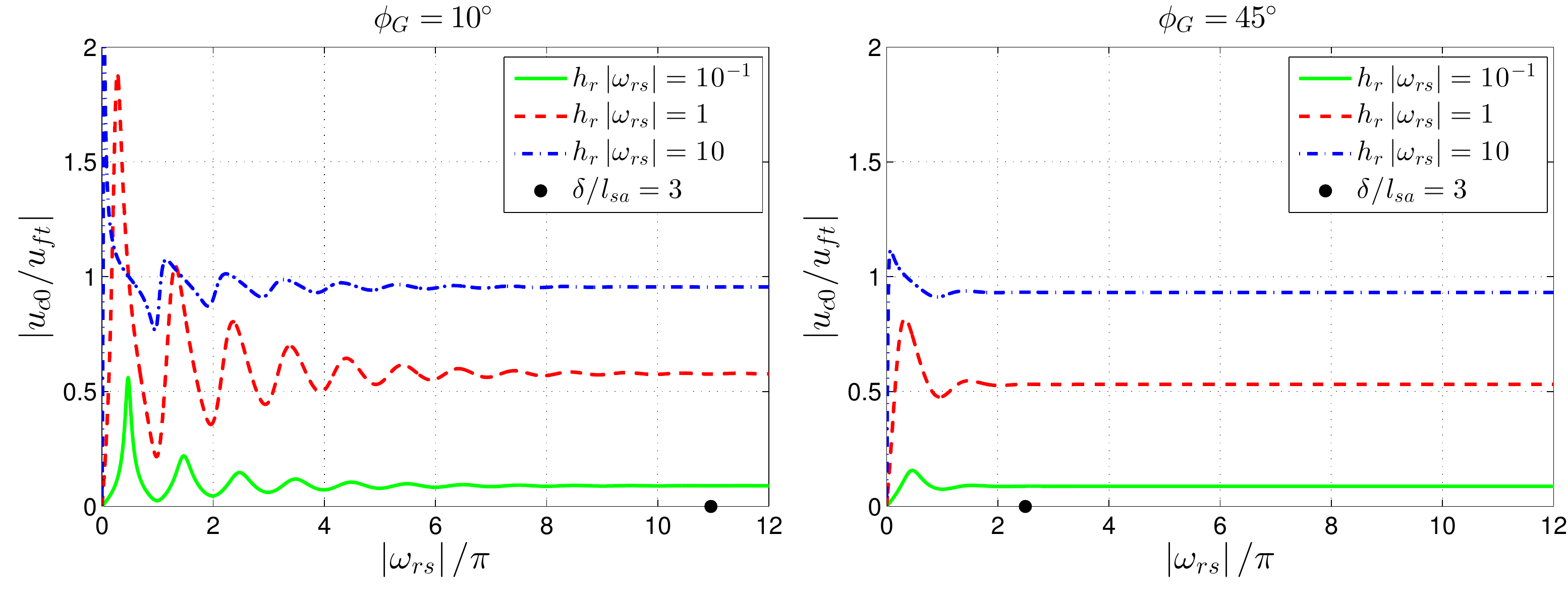}
\end{minipage}
\caption{Magnitude of the normalized interface velocity as function of $\left| \omegars \right|$ for three different values of $\hr \left| \omegars \right|$ and $\rhor = 1$, $\omegarf = 1$. 
The increase of $\left| \omegars \right|$ results from an increase of the coating thickness.
Left: low damping with $\phiG = 10\degree$. Right: high damping with $\phiG = 45\degree$.
}
\label{fig:interface_velocity_different_hr_omegars}
\end{figure}

Figure \ref{fig:interface_velocity_different_omegarf} shows the magnitude of the normalized surface velocity for two \textit{viscoelastic} solids: one with low damping ($\phiG = 10\degree$) and another with high damping ($\phiG = 45\degree$). The resonances are less strong or even absent. For $\phiG = 10\degree$, there are still minima in the interface velocity, but no-slip does not occur any more for non-zero $\omegars$. In addition, not every $\omegarf$ allows for a solution with $\abs{\usurfnot/\uft} = 1$, which contrasts with the purely elastic case. The two subfigures also hint at the limiting cases that have been discussed in section \ref{sec:dynamics_limiting_cases}. When $\omegarf \goto \infty$, the surface velocity becomes zero (compare with figure \ref{fig:velocity_profiles_different_omegarf}). No-slip occurs for $\abs{\omegars} \goto 0$ (a stiff solid) and free-slip for $\abs{\omegars} \goto \infty$ (a soft solid).

The influence of coating thickness is shown in figure \ref{fig:interface_velocity_different_hr_omegars}. It displays the magnitude of the normalized interface velocity as function of $\abs{\omegars}$ at fixed $\hr \abs{\omegars} = \omega \halfc / \abs{\cs}$. The increase of $\abs{\omegars}$ on the horizontal axis results from an increase of the coating thickness. The graphs display some oscillations at low $\abs{\omegars}$, like in figure \ref{fig:interface_velocity_different_omegarf}. However, the interface velocity becomes constant above a certain $\abs{\omegars}$. This can be explained with use of the decay length $\lsa = 1/\ksa$ introduced below equation \ref{eq:superposition_decaying_travelling_waves}. The exponential function decays to 5\% of its initial amplitude in 3 decay lengths. Hence, $\usurf$ will be independent of $\tc$ when $\tc \gtrsim 3 \lsa$ or $\tc/\lsa = \abs{\omegars} \sin{\phase{\cs}} \gtrsim 3$. Figure \ref{fig:interface_velocity_different_hr_omegars} indicates with a solid circle the value of $\abs{\omegars}$ for which $\tc / \lsa = 3$. Indeed, the graphs confirm that the interface velocity becomes independent of coating thickness when $\tc \gtrsim 3 \lsa$. In that case, the relevant dimensionless parameter becomes $\hr \, \omegars = \omega \halfc / \cs$, which is $\omegars$ with $\tc$ replaced by $\halfc$.

\begin{figure}
\centering
\begin{minipage}[t]{1\linewidth}
\centering
\includegraphics[width=\textwidth]{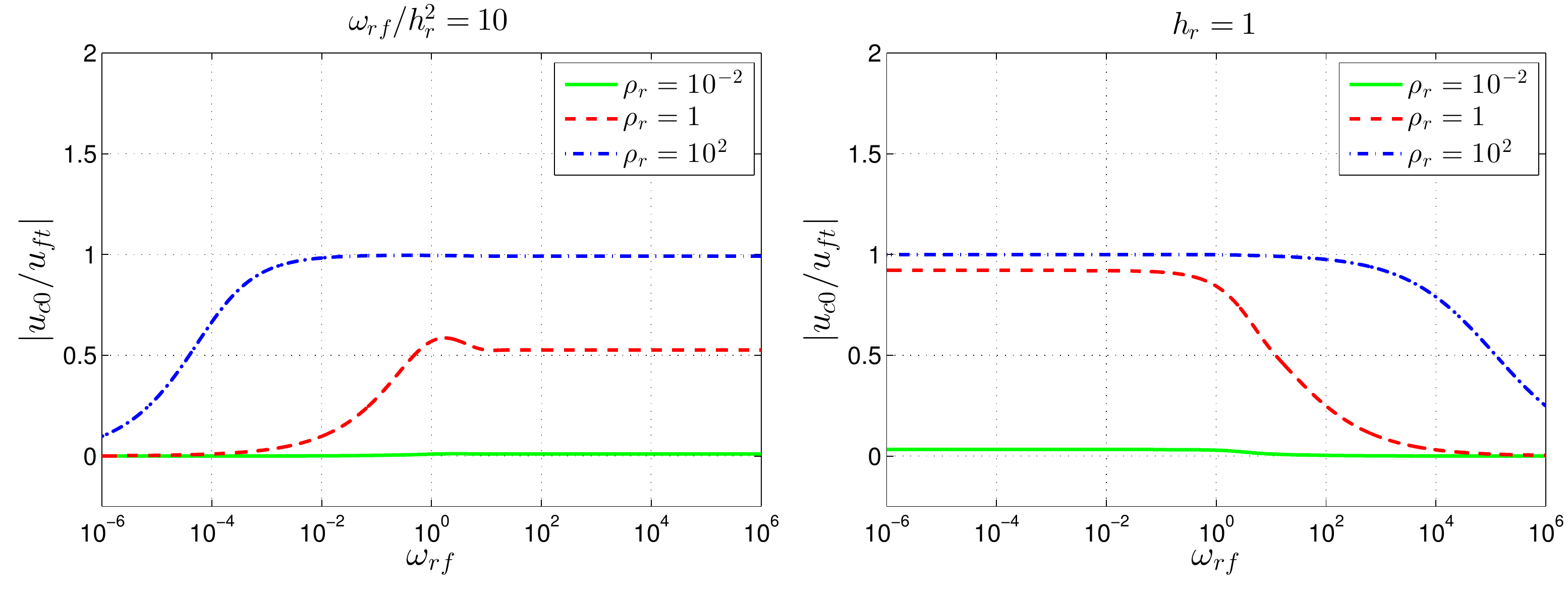}
\end{minipage}
\caption{Magnitude of the normalized interface velocity as function of $\omegarf$ for three different values of $\rhor$ and $\left| \omegars \right| = 2\pi$, $\phiG = 10\degree$. 
Left: the increase of $\omegarf$ results from an increase of the half-channel height; $\omegarf / \hr^2 = \omega \tc^2 / \nu = 10$. 
Right: the increase of $\omegarf$ results from a decrease of the kinematic viscosity; $\hr = 1$.
}
\label{fig:interface_velocity_different_rhor}
\end{figure}

The influence of the half-channel height $\halfc$ is shown in figure \ref{fig:interface_velocity_different_rhor} (left). It displays the magnitude of the normalized interface velocity as function of $\omegarf$ at fixed $\omegarf/\hr^2 = \omega \tc^2 / \nu$. The increase of $\omegarf$ on the horizontal axis results from an increase of the half-channel height at fixed frequency and kinematic viscosity. When $\halfc \goto 0$, both $\omegarf \goto 0$ and $\hr \goto 0$, which results in $\abs{\intpar} \goto 0$ (equation \ref{eq:intpar_small_omegarf}) and $\abs{\usurfnot} \goto 0$. However, when $\halfc \goto \infty$, the interface velocity approaches a constant value for $\omegarf \gtrsim 10^2$. Indeed, $\intpar$ is independent of $\halfc$ when $\omegarf \goto \infty$ (equation \ref{eq:intpar_large_omegarf}). Physically, the channel height is so large that it is irrelevant for the interface dynamics. The Stokes length  $\lstokes = \sqrt{ 2\nu/\omega }$ is the only important length scale for the fluid. The relevant dimensionless number becomes $\omegarf / \hr^2 = \omega \tc^2 / \nu$, which is $\omegarf$ with $\halfc$ replaced by $\tc$ (compare with equation \ref{eq:intpar_large_omegarf}).

The influence of the kinematic viscosity $\nu$ is shown in figure \ref{fig:interface_velocity_different_rhor} (right). It displays the magnitude of the normalized interface velocity as function of $\omegarf$ at fixed $\hr$. The increase of $\omegarf$ on the horizontal axis results from a decrease of the kinematic viscosity at fixed frequency and half-channel height. When $\nu \goto 0$, $\omegarf \goto \infty$ and $\abs{\intpar} \goto 0$ (equation \ref{eq:intpar_large_omegarf}), which yields $\abs{\usurfnot} \goto 0$. This has been already observed in figure \ref{fig:velocity_profiles_different_omegarf} (right). The shear stresses become too small to deform the coating. 
However, when $\nu \goto \infty$, the interface velocity approaches a constant value for $\omegarf \lesssim 10^{-2}$. Indeed, $\intpar$ is independent of $\nu$ when $\omegarf \goto 0$ (equation \ref{eq:intpar_small_omegarf}). The reason is that the shear stress becomes independent of the kinematic viscosity for quasi-steady flow, as is proven in the next subsection. 

Finally, figure \ref{fig:interface_velocity_different_rhor} also shows the dependence on the density ratio $\rhor$. The no-slip condition with $\usurfnot = 0$ is obtained in the limit $\rhor \goto 0$. On the other hand, the free-slip condition with $\abs{\usurfnot / \uft} = 1$ is obtained when $\rhor \goto \infty$. These trends have been already observed in figure \ref{fig:three_types_global_dynamics_density_ratio}.




\subsection{Interface shear stress}

The wall shear stress $\tauw$ is a relevant parameter in the field of drag reduction. It is defined by:
\begin{subequations}
\label{eq:interface_shear_stress}
\begin{align}
\tauw & = \mu \left. \! \pd{\uf}{z} \right|_{z = -\halfc} = \tauwnot \expp{\comp \omega t}, \\ 
\tauwt & = \rhof \fnot \halfc = 
\mu \, \frac{\fnot \halfc^2 / \nu}{\halfc} = 
\sqrt{\omegarf} \cdot \mu \, \frac{\uft}{\sqrt{\nu/\omega}}, \\
\tauwr & = \frac{\tauwnot}{\tauwt} = \lb \frac{1}{1-\intpar} \rb \frac{ \tanh{\lb \sqrt{\comp \omegarf} \rb} }{ \sqrt{\comp \omegarf} },
\end{align}
\end{subequations}
where $\tauwnot$ is the shear stress amplitude, $\tauwt$ is a typical shear stress and $\tauwr$ is the shear stress amplitude relative to this typical shear stress. 
Note that $\tauwt$ represents a typical fluid shear stress when the solid is rigid ($\intpar = 0$); it only depends on fluid and forcing parameters. However, the choice for $\tauwt$ is not unique. The present choice corresponds with the shear stress magnitude in the limit $\omegarf \goto 0$ and $\intpar = 0$: equation \ref{eq:interface_shear_stress} then gives $\tauwnot = \rhof \fnot \halfc = \mu (\fnot \halfc^2 / \nu) / \halfc$. The wall shear stress is balanced by the forcing in this quasi-steady case. The relevant scales are $\fnot \halfc^2 / \nu$ for velocity and $\halfc$ for length; $\omega$ disappears.
Another choice could be the shear stress in the limit $\omegarf \goto \infty$ (and $\intpar = 0$). Equation \ref{eq:interface_shear_stress} then yields $\abs{\tauwnot} = \rhof \fnot \halfc / \sqrt{\omegarf} = \mu \, \uft/\sqrt{\nu/\omega}$ as typical shear stress. The relevant scales are $\uft$ for velocity and $\sqrt{\nu/\omega}$ for length; $\halfc$ disappears.
However, we prefer a frequency-independent $\tauwt$, because $\omega$ already appears in the dimensionless numbers $\omegars$ and $\omegarf$. Therefore, $\rhof \fnot \halfc$ has been chosen as a typical shear stress.


\begin{figure}
\centering
\begin{minipage}[t]{1\linewidth}
\centering
\includegraphics[width=\textwidth]{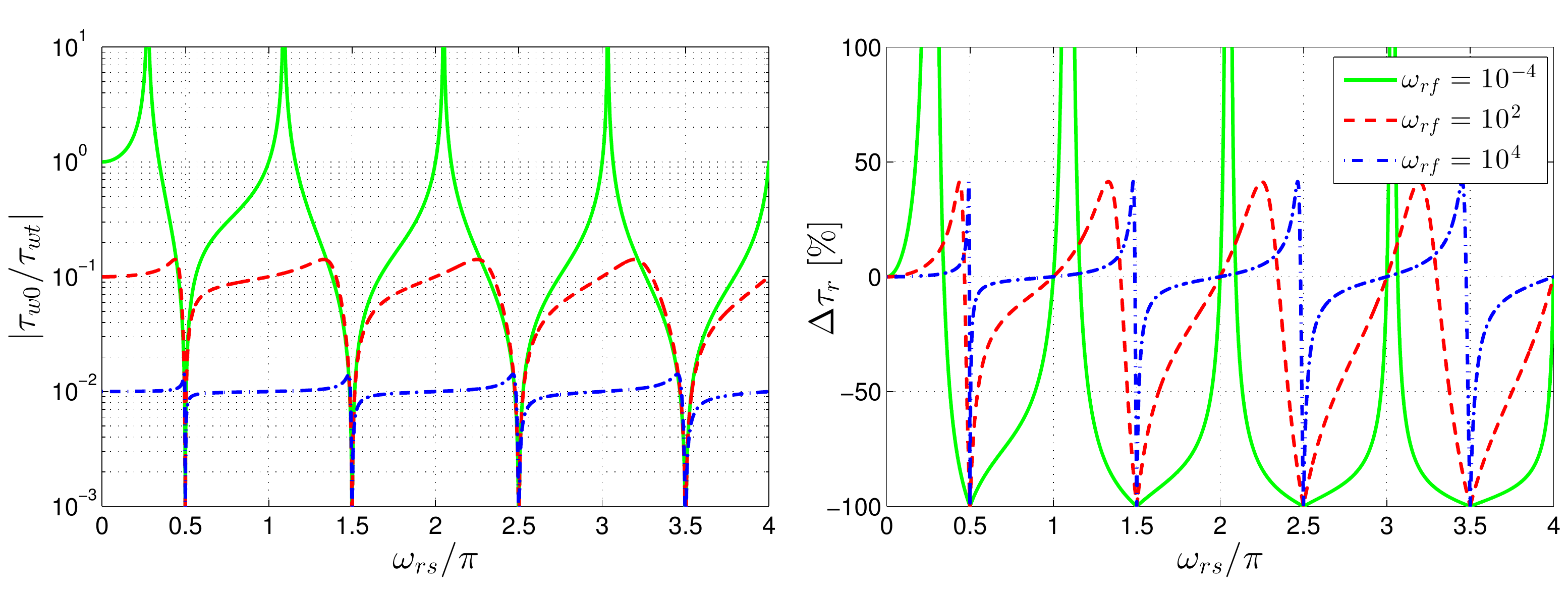}
\end{minipage}
\caption{Quantities related to the shear stress magnitude as function of $\omegars$ for three different values of $\omegarf$ and $\rhor = 1$, $\hr = 1$, $\phiG = 0\degree$. 
The increase of $\omegars$ results from a decrease of the coating stiffness.
Left: the shear stress magnitude relative to a typical shear stress. 
Right: the drag change, which quantifies the change of the shear stress magnitude relative to a rigid wall.
}
\label{fig:shear_stress_amplitude}
\end{figure}

\begin{figure}[t!]
\centering
\begin{minipage}[t]{1\linewidth}
\centering
\includegraphics[width=\textwidth]{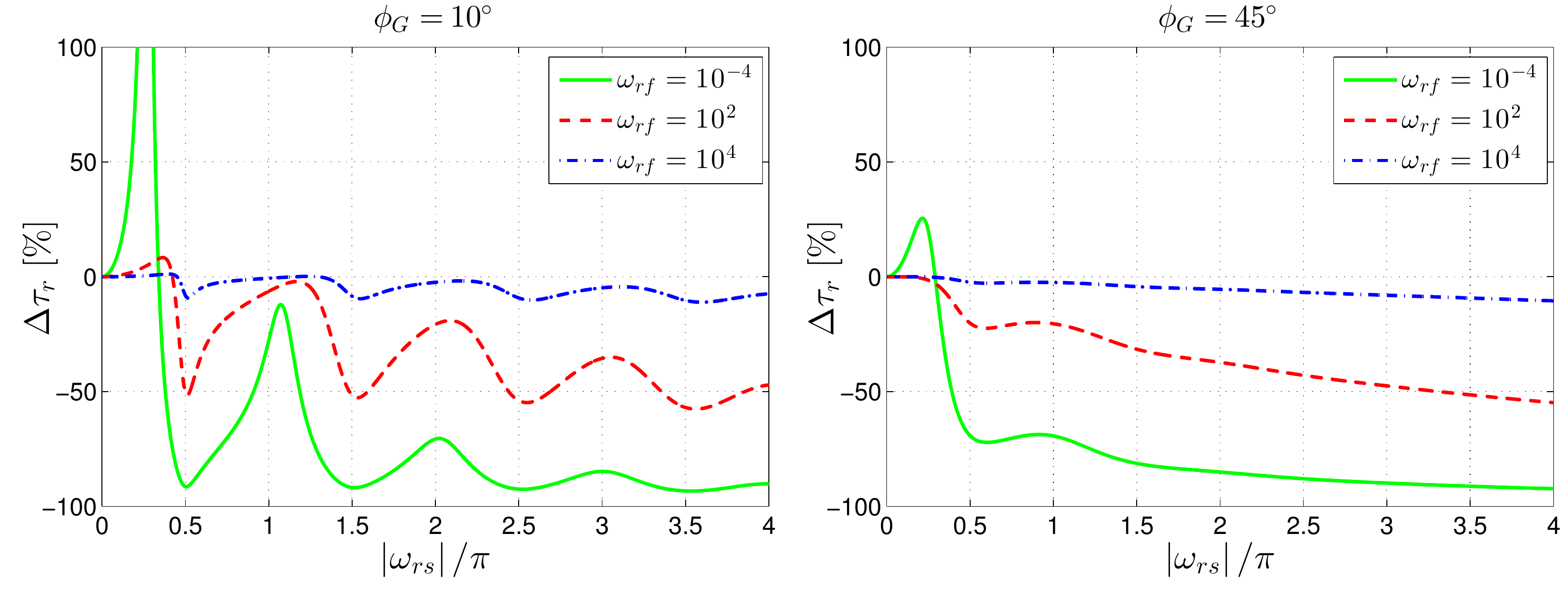}
\end{minipage}
\caption{The drag change as function of $\left| \omegars \right|$ for three different values of $\omegarf$ and $\rhor = 1$, $\hr = 1$. 
The increase of $\left| \omegars \right|$ results from a decrease of the coating stiffness.
Left: low damping with $\phase{\shearmod} = 10\degree$. 
Right: high damping with $\phase{\shearmod} = 45\degree$.
}
\label{fig:drag_change_different_damping}
\end{figure}

In the context of drag reduction, it is natural to investigate how the shear stress magnitude changes due to the presence of the coating. Therefore, the drag change $\deltataur$ is introduced:
\begin{equation}
\begin{aligned}
\deltataur = \frac{ \abs{\tauwr} - \abs{\tauwrrigid} }{ \abs{\tauwrrigid} }
= \frac{1}{\, \abs{1 - \intpar} \,} - 1,
\end{aligned}
\end{equation}
where $\tauwrrigid$ is the relative shear stress amplitude when the wall is rigid (so $\intpar = 0$). The parameter $\deltataur$ quantifies the change of the shear stress magnitude relative to a rigid wall.
Note that $\deltataur$ depends exclusively on $\intpar$, like $\usurfnot/\uft$. Hence, most findings related to the interface velocity also apply to the shear stress change. Therefore, this subsection will be short; only two figures are included.

Figure \ref{fig:shear_stress_amplitude} shows the normalized shear stress magnitude and the shear stress change as function of $\omegars$ for an \textit{elastic} solid and three different values of $\omegarf$. The figure exhibits many features that have already appeared in figure \ref{fig:interface_velocity_amplitude_and_phase}, such as the even, odd and coupling resonances. For the even modes ($\abs{\intpar} \goto 0$), the shear stress is the same as for a rigid wall and $\deltataur = 0$. For the odd modes ($\abs{\intpar} \goto \infty$), the fluid velocity is uniform, the shear stress equals zero, and 100\% drag reduction is obtained. For the coupling modes ($\real{\intpar} \approx 1$), the large interface velocity is accompanied by a maximum drag increase.

Figure \ref{fig:drag_change_different_damping} shows the drag change for a \textit{viscoelastic} solid. The coupling resonances are less strong or have disappeared completely. The odd and even resonance modes are also affected. The even modes do not correspond with $\deltataur = 0$ and the odd modes do not reach 100\% drag reduction any more. However, the benefit of viscoelasticity is that the response of the fluid-solid system changes less dramatically near the resonances.

\begin{figure}
\centering
\begin{minipage}[t]{1\linewidth}
\centering
\includegraphics[width=\textwidth]{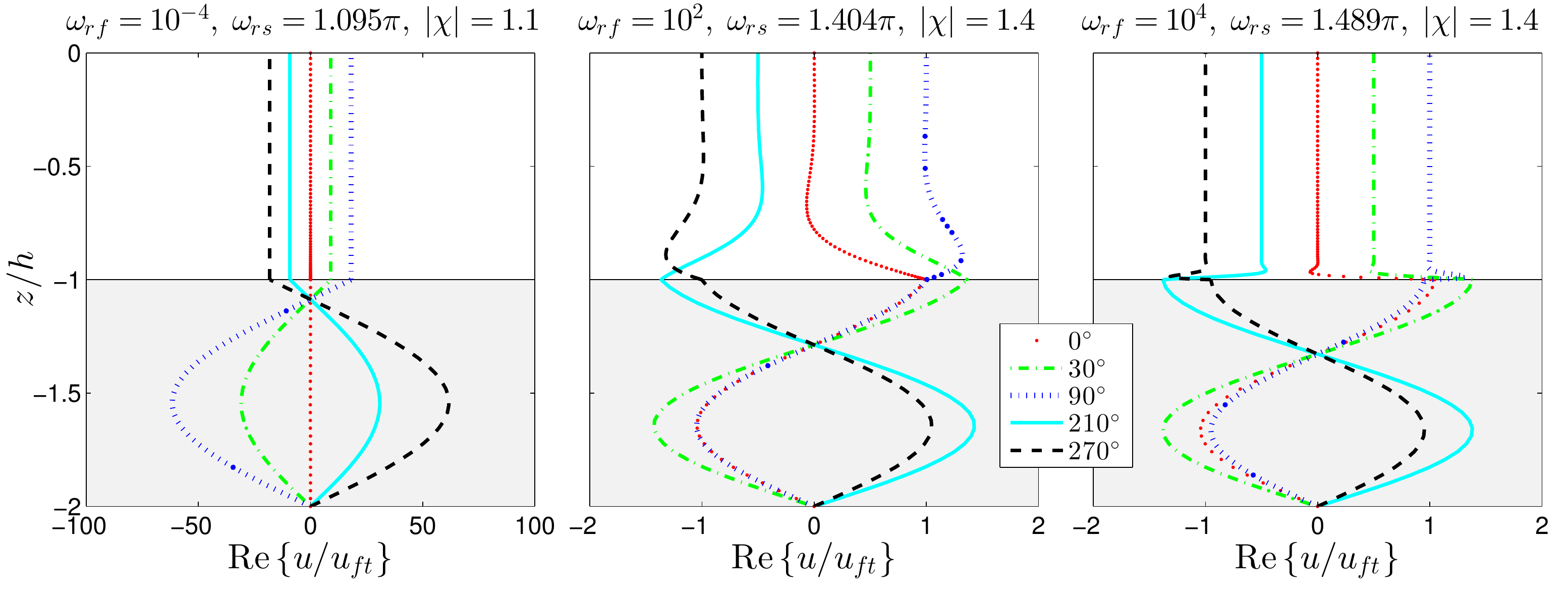}
\end{minipage}
\caption{Velocity profiles of coupling modes for three different values of $\omegarf$ and $\omegars$ at five different phase angles. The other parameters are $\rhor = 1$, $\hr = 1$, $\phiG = 0\degree$.
}
\label{fig:velocity_profiles_coupling_modes}
\end{figure}

\subsection{Resonances}

The previous two subsections have shown that three types of resonances can occur, namely odd, even and coupling modes. The odd and even modes are specific to the solid. Figure \ref{fig:velocity_profiles_different_omegars} has already shown examples of velocity profiles for odd and even modes. Figure \ref{fig:velocity_profiles_coupling_modes} shows velocity profiles for coupling modes in an elastic solid at three different $\omegarf$. The amplitude of the normalized interface velocity is indeed larger than one: $\abs{\usurfnot/\uft}>1$, in agreement with figure \ref{fig:interface_velocity_amplitude_and_phase}. The velocity magnitude is especially large for low $\omegarf$.



Near resonances, the response of the fluid-solid system changes dramatically, as is also observed in other studies \citep{luhar2016design}. For instance, a large drag reduction might become a large drag increase when $\omegars$ is changed slightly, especially for elastic solids and low $\omegarf$ (see figure \ref{fig:shear_stress_amplitude}). Such behaviour might not be desirable for applications. In that case, one can tune the parameters such that resonances are less strong or do not occur at all. For instance, one could require that $\omegars \lesssim 1$ for an elastic solid. As a second example, one could add damping to the solid. Resonances are less strong when the solid is more viscous \citep{kulik2008wave}. That is also evident from the comparison of figures \ref{fig:interface_velocity_amplitude_and_phase} and \ref{fig:interface_velocity_different_omegarf}, or figures \ref{fig:shear_stress_amplitude} and \ref{fig:drag_change_different_damping}.

\subsection{Compliant coatings in practice}

The purpose of this subsection is to investigate when deformation of compliant coatings by fluid shear is significant in practice. Consider an FSI problem with fixed forcing frequency, channel geometry and solid material properties. Specifically, the following parameters have been chosen: $\omega / (2 \pi) = 10$ Hz, $h = 1$ cm, $\rhos = 1000 \un{kg\,m^{-3}}$, $\abs{\shearmod} = 300$ Pa and $\phiG = 10\degree$. The coating is extremely soft \citep{stirling2015novel}. Three different fluids and two different coating thicknesses are considered, such that six possible combinations result (see table \ref{tab:dimensionless_numbers_practice}). The three fluids are air, water and a viscous oil, all at room temperature. As the fluids have different densities and viscosities, they influence the interaction through $\rhor$ and $\omegarf$. The coating thicknesses are 1 mm (thin) and 10 cm (thick), which results in two different values of $\omegars$ at fixed $\hr \, \omegars$. The coating of 10 cm, although very thick, is not yet so thick that the interface velocity becomes independent of the coating thickness, because $\tc / \lsa = 1.0 < 3$ (compare with figure \ref{fig:interface_velocity_different_hr_omegars}).


Table \ref{tab:dimensionless_numbers_practice} shows the fluid and coating parameters with the corresponding dimensionless numbers, including the magnitude of the shear interaction parameter and the normalized interface velocity. An increase of $\abs{\intpar}$ and $\abs{\usurfnot / \uft}$ indicates a more significant interaction between fluid and solid.
Air has the smallest interaction. Although it has a kinematic viscosity which is 15 times larger than that of water, its density is smaller by a factor of about 1000. The low density of air is the main reason for its small interaction parameter.
Water has a larger $\abs{\intpar}$, mainly because of its larger density.
By far the largest interaction is obtained with the viscous oil. Its density is comparable to that of water, but it is more viscous by a factor of 3000. That yields a much lower $\omegarf$ and a more significant interaction, as has been already observed from figures \ref{fig:velocity_profiles_different_omegarf}, \ref{fig:interface_velocity_different_omegarf} and \ref{fig:interface_velocity_different_rhor}. The table also shows that a thicker coating gives a larger interaction because of the $\abs{\omegars}$ increase.
In summary, significant interactions between a compliant coating and laminar shear flow are typically obtained for viscous, heavy fluids and soft, thick coatings.

\renewcommand{\arraystretch}{1.2}
\begin{table*}
  \centering
  \begin{minipage}[t]{1\linewidth}
  \caption{Fluid and coating parameters that could represent realistic experiments, with the corresponding dimensionless numbers. Fixed parameters are $\omega / (2 \pi) = 10$ Hz, $h = 1$ cm, $\rhos = 1000 \un{kg\,m^{-3}}$, $\left| \shearmod \right| = 300$ Pa and $\phiG = 10\degree$. Three fluids (air, water and viscous oil) and two coating thicknesses (thin and thick) are considered.}
  \resizebox{\columnwidth}{!}{
    \begin{tabular}{ l l l l l l l l l l l l l }
    \hline
Description	 & $\rhof$	 & $\nu$	 & $\delta$	 & $\rhor$	 & $\hr$	 & $\omegarf$	 & $\left| \omegars \right| / \pi$	 & $\left| \intpar \right|$	 & $\left| \usurfnot / \uft \right|$ \\ 
 	 & \unit{kg\,m^{-3}}	 & \unit{m^2s^{-1}}	 & \unit{m}	 &  	 &  	 &  	 &  	 &  	 &   \\ 
 	 \hline
	Air, thin	 & 1.2	 & $1.5 \cdot 10^{-5}$	 & 0.001	 & 0.0012	 & 10	 & $4.2 \cdot 10^{2}$	 & 0.037	 & $7.8 \cdot 10^{-6}$	 & $7.8 \cdot 10^{-6}$ \\ 
	Air, thick	 & 1.2	 & $1.5 \cdot 10^{-5}$	 & 0.1	 & 0.0012	 & 0.1	 & $4.2 \cdot 10^{2}$	 & 3.7	 & $8.0 \cdot 10^{-5}$	 & $8.0 \cdot 10^{-5}$ \\ 
	Water, thin	 & 1000	 & $1.0 \cdot 10^{-6}$	 & 0.001	 & 1	 & 10	 & $6.3 \cdot 10^{3}$	 & 0.037	 & $1.7 \cdot 10^{-3}$	 & $1.7 \cdot 10^{-3}$ \\ 
	Water, thick	 & 1000	 & $1.0 \cdot 10^{-6}$	 & 0.1	 & 1	 & 0.1	 & $6.3 \cdot 10^{3}$	 & 3.7	 & $1.7 \cdot 10^{-2}$	 & $1.7 \cdot 10^{-2}$ \\ 
	Oil, thin	 & 950	 & $3.0 \cdot 10^{-3}$	 & 0.001	 & 0.95	 & 10	 & 2.1	 & 0.037	 & $9.8 \cdot 10^{-2}$	 & $1.0 \cdot 10^{-1}$ \\ 
	Oil, thick	 & 950	 & $3.0 \cdot 10^{-3}$	 & 0.1	 & 0.95	 & 0.1	 & 2.1	 & 3.7	 & 1.0	 & $5.4 \cdot 10^{-1}$ \\ 
	\hline
    \end{tabular}
    }
  \label{tab:dimensionless_numbers_practice}%
  \end{minipage}
\end{table*}

\section{Conclusions \& Discussion} \label{sec:discussion}

This paper has investigated analytically a fundamental problem in fluid-structure interaction: the oscillatory laminar shear flow over a compliant viscoelastic layer on a rigid base. 
An analytical solution of the fluid and solid velocity has been obtained. Five dimensionless parameters appear. However, the interaction between fluid and solid is determined by one combined dimensionless parameter, which we call the shear interaction parameter $\intpar$. The fluid satisfies a no-slip boundary condition when $\abs{\intpar} \goto 0$, which occurs when the solid is heavy, stiff and/or thin. In contrast, the fluid obeys free-slip when $\abs{\intpar} \goto \infty$, which corresponds to a lightweight and/or soft solid.
The dynamics of the fluid-solid system is characterized by resonant behaviour when the solid is elastic. Three types of resonance modes have been identified. Two modes (odd and even) are specific to the solid. The third mode results from the coupling with the fluid, which yields an enlarged velocity and shear stress at the interface.
A viscoelastic solid damps the oscillations. The three resonance modes are less strong or even absent. The shear stress at the interface is typically reduced.

The findings in this paper have a twofold use. 
First, they help to understand the fluid and solid dynamics when shear coupling is important, such as in wall-bounded shear flows.
Second, the presented analytical solution is very useful for validation of numerical FSI solvers. According to \citet{gad2002compliant}, any numerical result of a FSI problem is - without sufficient validation - suspect at best and wrong at worst.

The main limitation of the present study relates to stability. The derived analytical solution is only valid as long as the flow and solid are stable. That requires a sufficiently low Reynolds number and a sufficiently stiff coating. Above a critical Reynolds number, the laminar flow will become transitional and eventually turbulent, which involves more complex interactions. Regarding coating stiffness, the current work suggests that 100\% reduction of shear stress can be obtained with very soft coatings (so $\abs{\omegars} \goto \infty$, see e.g. figure \ref{fig:drag_change_different_damping}). However, soft coatings are also susceptible to flow-induced surface instabilities, such as travelling surface waves \citep{duncan1985dynamics, gad2002compliant}. A stability analysis, although beyond the scope of the current paper, might clarify when the interaction between the oscillatory shear flow and the compliant layer becomes unstable.

The present work opens up a few interesting future directions. First, blood vessel walls are composed of three layers \citep{canic2014fluid}. The current problem can be extended to multiple solid layers with different viscoelastic properties. An analytical solution can presumably also be obtained for that case.
Second, normal stresses are important in both blood flows and turbulent flows. It is therefore natural to continue the present work by incorporation of the normal stresses.
Finally, there is still a need for Direct Numerical Simulations of turbulent flow over a viscoelastic layer \citep{kulik2008wave}. The analytical solution presented here can be very useful to validate such simulations. \\

The research leading to these results has received funding from the European Union Seventh Framework Programme in the SEAFRONT project under grant agreement nr. 614034.
%
%


\begin{thebibliography}{45}
\providecommand{\natexlab}[1]{#1}
\providecommand{\url}[1]{\texttt{#1}}
\expandafter\ifx\csname urlstyle\endcsname\relax
  \providecommand{\doi}[1]{doi: #1}\else
  \providecommand{\doi}{doi: \begingroup \urlstyle{rm}\Url}\fi

\bibitem[Bergel(1961)]{bergel1961dynamic}
D.~Bergel.
\newblock The dynamic elastic properties of the arterial wall.
\newblock \emph{The Journal of Physiology}, 156\penalty0 (3):\penalty0 458,
  1961.

\bibitem[{\v{C}}ani{\'c} et~al.(2014){\v{C}}ani{\'c}, Muha, and
  Buka{\v{c}}]{canic2014fluid}
S.~{\v{C}}ani{\'c}, B.~Muha, and M.~Buka{\v{c}}.
\newblock {Fluid--structure interaction in hemodynamics: Modeling, analysis,
  and numerical simulation}.
\newblock In \emph{Fluid-Structure Interaction and Biomedical Applications},
  pages 79--195. Springer, 2014.

\bibitem[Carcione(2015)]{carcione2015wave}
J.~M. Carcione.
\newblock \emph{Wave Fields in Real Media (Third Edition)}.
\newblock 2015.

\bibitem[Carfagni et~al.(1998)Carfagni, Lenzi, and Pierini]{carfagni1998loss}
M.~Carfagni, E.~Lenzi, and M.~Pierini.
\newblock The loss factor as a measure of mechanical damping.
\newblock In \emph{Proceedings of the 16th International Modal Analysis
  Conference}, volume 3243, page 580, 1998.

\bibitem[Carpenter and Garrad(1986)]{carpenter1986hydrodynamic}
P.~Carpenter and A.~Garrad.
\newblock {The hydrodynamic stability of flow over Kramer-type compliant
  surfaces. Part 2. Flow-induced surface instabilities}.
\newblock \emph{Journal of Fluid Mechanics}, 170:\penalty0 199--232, 1986.

\bibitem[Choi et~al.(1997)Choi, Yang, Clayton, Glover, Atlar, Semenov, and
  Kulik]{choi1997}
K.-S. Choi, X.~Yang, B.~R. Clayton, E.~J. Glover, M.~Atlar, B.~N. Semenov, and
  V.~M. Kulik.
\newblock Turbulent drag reduction using compliant surfaces.
\newblock \emph{Proceedings of the Royal Society of London. Series A:
  Mathematical, Physical and Engineering Sciences}, 453\penalty0
  (1965):\penalty0 2229--2240, 1997.

\bibitem[Christensen(1982)]{christensen1982theory}
R.~Christensen.
\newblock \emph{Theory of Viscoelasticity (Second Edition)}.
\newblock 1982.

\bibitem[Chung(2001)]{chung2001review}
D.~Chung.
\newblock {Review: Materials for vibration damping}.
\newblock \emph{Journal of Materials Science}, 36\penalty0 (24):\penalty0
  5733--5737, 2001.

\bibitem[Duncan et~al.(1985)Duncan, Waxman, and Tulin]{duncan1985dynamics}
J.~Duncan, A.~Waxman, and M.~Tulin.
\newblock The dynamics of waves at the interface between a viscoelastic coating
  and a fluid flow.
\newblock \emph{Journal of Fluid Mechanics}, 158:\penalty0 177--197, 1985.

\bibitem[Duncan(1986)]{duncan1986response}
J.~H. Duncan.
\newblock The response of an incompressible, viscoelastic coating to pressure
  fluctuations in a turbulent boundary layer.
\newblock \emph{Journal of Fluid Mechanics}, 171:\penalty0 339--363, 1986.

\bibitem[Endo and Himeno(2002)]{endo2002direct}
T.~Endo and R.~Himeno.
\newblock Direct numerical simulation of turbulent flow over a compliant
  surface.
\newblock \emph{Journal of Turbulence}, 3:\penalty0 7, 2002.

\bibitem[Fabre et~al.(2012)Fabre, Gilbert, Hirschberg, and
  Pelorson]{fabre2012aeroacoustics}
B.~Fabre, J.~Gilbert, A.~Hirschberg, and X.~Pelorson.
\newblock Aeroacoustics of musical instruments.
\newblock \emph{Annual Review of Fluid Mechanics}, 44:\penalty0 1--25, 2012.

\bibitem[Fukagata et~al.(2008)Fukagata, Kern, Chatelain, Koumoutsakos, and
  Kasagi]{fukagata2008evolutionary}
K.~Fukagata, S.~Kern, P.~Chatelain, P.~Koumoutsakos, and N.~Kasagi.
\newblock Evolutionary optimization of an anisotropic compliant surface for
  turbulent friction drag reduction.
\newblock \emph{Journal of Turbulence}, \penalty0 (9):\penalty0 N35, 2008.

\bibitem[Gad-el Hak(2002)]{gad2002compliant}
M.~Gad-el Hak.
\newblock Compliant coatings for drag reduction.
\newblock \emph{Progress in Aerospace Sciences}, 38\penalty0 (1):\penalty0
  77--99, 2002.

\bibitem[Gad-El-Hak et~al.(1984)Gad-El-Hak, Blackwelder, and
  Riley]{gad1984interaction}
M.~Gad-El-Hak, R.~F. Blackwelder, and J.~J. Riley.
\newblock On the interaction of compliant coatings with boundary-layer flows.
\newblock \emph{Journal of Fluid Mechanics}, 140:\penalty0 257--280, 1984.

\bibitem[Grotberg and Jensen(2004)]{grotberg2004biofluid}
J.~B. Grotberg and O.~E. Jensen.
\newblock Biofluid mechanics in flexible tubes.
\newblock \emph{Annual Review of Fluid Mechanics}, 36:\penalty0 121--147, 2004.

\bibitem[Gundogdu and Carpinlioglu(1999)]{gundogdu1999present1}
M.~Y. Gundogdu and M.~O. Carpinlioglu.
\newblock Present state of art on pulsatile flow theory. part 1. laminar and
  transitional flow regimes.
\newblock \emph{JSME International Journal Series B Fluids and Thermal
  Engineering}, 42\penalty0 (3):\penalty0 384--397, 1999.

\bibitem[Hale et~al.(1955)Hale, McDonald, and Womersley]{hale1955velocity}
J.~Hale, D.~McDonald, and J.~Womersley.
\newblock Velocity profiles of oscillating arterial flow, with some
  calculations of viscous drag and the reynolds number.
\newblock \emph{The Journal of Physiology}, 128\penalty0 (3):\penalty0 629,
  1955.

\bibitem[Hamadiche and Gad-el Hak(2004)]{hamadiche2004spatiotemporal}
M.~Hamadiche and M.~Gad-el Hak.
\newblock Spatiotemporal stability of flow through collapsible, viscoelastic
  tubes.
\newblock \emph{AIAA Journal}, 42\penalty0 (4):\penalty0 772--786, 2004.

\bibitem[Heil and Hazel(2011)]{heil2011fluid}
M.~Heil and A.~L. Hazel.
\newblock Fluid-structure interaction in internal physiological flows.
\newblock \emph{Annual Review of Fluid Mechanics}, 43:\penalty0 141--162, 2011.

\bibitem[Kamakoti and Shyy(2004)]{kamakoti2004fluid}
R.~Kamakoti and W.~Shyy.
\newblock Fluid--structure interaction for aeroelastic applications.
\newblock \emph{Progress in Aerospace Sciences}, 40\penalty0 (8):\penalty0
  535--558, 2004.

\bibitem[Kim and Choi(2014)]{kim2014space}
E.~Kim and H.~Choi.
\newblock Space--time characteristics of a compliant wall in a turbulent
  channel flow.
\newblock \emph{Journal of Fluid Mechanics}, 756:\penalty0 30--53, 2014.

\bibitem[Kulik(2012)]{kulik2012action}
V.~Kulik.
\newblock Action of a turbulent flow on a hard compliant coating.
\newblock \emph{International Journal of Heat and Fluid Flow}, 33\penalty0
  (1):\penalty0 232--241, 2012.

\bibitem[Kulik et~al.(2008)Kulik, Lee, and Chun]{kulik2008wave}
V.~M. Kulik, I.~Lee, and H.~Chun.
\newblock Wave properties of coating for skin friction reduction.
\newblock \emph{Physics of Fluids (1994-present)}, 20\penalty0 (7):\penalty0
  075109, 2008.

\bibitem[Kumaran(1995)]{kumaran1995stability}
V.~Kumaran.
\newblock Stability of the viscous flow of a fluid through a flexible tube.
\newblock \emph{Journal of Fluid Mechanics}, 294:\penalty0 259--281, 1995.

\bibitem[Lautrup(2011)]{lautrup2011physics}
B.~Lautrup.
\newblock \emph{Physics of Continuous Matter, Second Edition: Exotic and
  Everyday Phenomena in the Macroscopic World}.
\newblock Taylor \& Francis, 2011.

\bibitem[Lee et~al.(1993)Lee, Fisher, and Schwarz]{lee1993investigation}
T.~Lee, M.~Fisher, and W.~Schwarz.
\newblock Investigation of the stable interaction of a passive compliant
  surface with a turbulent boundary layer.
\newblock \emph{Journal of Fluid Mechanics}, 257:\penalty0 373--401, 1993.

\bibitem[Luhar et~al.(2015)Luhar, Sharma, and McKeon]{luhar2015framework}
M.~Luhar, A.~S. Sharma, and B.~McKeon.
\newblock A framework for studying the effect of compliant surfaces on wall
  turbulence.
\newblock \emph{Journal of Fluid Mechanics}, 768:\penalty0 415--441, 2015.

\bibitem[Luhar et~al.(2016)Luhar, Sharma, and McKeon]{luhar2016design}
M.~Luhar, A.~Sharma, and B.~McKeon.
\newblock On the design of optimal compliant walls for turbulence control.
\newblock \emph{Journal of Turbulence}, 17\penalty0 (8):\penalty0 787--806,
  2016.

\bibitem[Nakamura and Kaneko(2008)]{nakamura2008flow}
T.~Nakamura and S.~Kaneko.
\newblock \emph{{Flow Induced Vibrations: Classifications and Lessons from
  Practical Experiences}}.
\newblock Elsevier, 2008.

\bibitem[Newman and Karniadakis(1997)]{newman1997direct}
D.~J. Newman and G.~E. Karniadakis.
\newblock A direct numerical simulation study of flow past a freely vibrating
  cable.
\newblock \emph{Journal of Fluid Mechanics}, 344:\penalty0 95--136, 1997.

\bibitem[Pandey et~al.(2016)Pandey, Karpitschka, Venner, and
  Snoeijer]{pandey2016lubrication}
A.~Pandey, S.~Karpitschka, C.~H. Venner, and J.~H. Snoeijer.
\newblock Lubrication of soft viscoelastic solids.
\newblock \emph{Journal of Fluid Mechanics}, 799:\penalty0 433--447, 2016.

\bibitem[Pipkin(1986)]{pipkin1986lectures}
A.~C. Pipkin.
\newblock \emph{Lectures on viscoelasticity theory}.
\newblock 1986.

\bibitem[Pluvinage et~al.(2014)Pluvinage, Kourta, and
  Bottaro]{pluvinage2014instabilities}
F.~Pluvinage, A.~Kourta, and A.~Bottaro.
\newblock Instabilities in the boundary layer over a permeable, compliant wall.
\newblock \emph{Physics of Fluids}, 26\penalty0 (8):\penalty0 084103, 2014.

\bibitem[Rebouillat and Liksonov(2010)]{rebouillat2010fluid}
S.~Rebouillat and D.~Liksonov.
\newblock {Fluid--structure interaction in partially filled liquid containers:
  A comparative review of numerical approaches}.
\newblock \emph{Computers \& Fluids}, 39\penalty0 (5):\penalty0 739--746, 2010.

\bibitem[Robertsson et~al.(1994)Robertsson, Blanch, and
  Symes]{robertsson1994viscoelastic}
J.~O. Robertsson, J.~O. Blanch, and W.~W. Symes.
\newblock Viscoelastic finite-difference modeling.
\newblock \emph{Geophysics}, 59\penalty0 (9):\penalty0 1444--1456, 1994.

\bibitem[Sotiropoulos et~al.(2016)Sotiropoulos, Le, and
  Gilmanov]{sotiropoulos2016fluid}
F.~Sotiropoulos, T.~B. Le, and A.~Gilmanov.
\newblock Fluid mechanics of heart valves and their replacements.
\newblock \emph{Annual Review of Fluid Mechanics}, 48:\penalty0 259--283, 2016.

\bibitem[Stirling and Zrinyi(2015)]{stirling2015novel}
T.~Stirling and M.~Zrinyi.
\newblock A novel method to determine the elastic modulus of extremely soft
  materials.
\newblock \emph{Soft Matter}, 11:\penalty0 4180--4188, 2015.

\bibitem[Tschoegl et~al.(2002)Tschoegl, Knauss, and Emri]{tschoegl2002poisson}
N.~W. Tschoegl, W.~G. Knauss, and I.~Emri.
\newblock Poisson's ratio in linear viscoelasticity--a critical review.
\newblock \emph{Mechanics of Time-Dependent Materials}, 6\penalty0
  (1):\penalty0 3--51, 2002.

\bibitem[Valdez-Jasso et~al.(2009)Valdez-Jasso, Haider, Banks, Santana,
  Germ{\'a}n, Armentano, and Olufsen]{valdez2009analysis}
D.~Valdez-Jasso, M.~A. Haider, H.~Banks, D.~B. Santana, Y.~Z. Germ{\'a}n, R.~L.
  Armentano, and M.~S. Olufsen.
\newblock Analysis of viscoelastic wall properties in ovine arteries.
\newblock \emph{IEEE Transactions on Biomedical Engineering}, 56\penalty0
  (2):\penalty0 210--219, 2009.

\bibitem[Vedeneev(2016)]{vedeneev2016propagation}
V.~Vedeneev.
\newblock Propagation of waves in a layer of a viscoelastic material underlying
  a layer of a moving fluid.
\newblock \emph{Journal of Applied Mathematics and Mechanics}, 80\penalty0
  (3):\penalty0 225--243, 2016.

\bibitem[Womersley(1955)]{womersley1955method}
J.~R. Womersley.
\newblock Method for the calculation of velocity, rate of flow and viscous drag
  in arteries when the pressure gradient is known.
\newblock \emph{The Journal of Physiology}, 127\penalty0 (3):\penalty0 553,
  1955.

\bibitem[Xie et~al.(2016)Xie, Zheng, Triantafyllou, Constantinides, and
  Karniadakis]{xie2016flow}
F.~Xie, X.~Zheng, M.~S. Triantafyllou, Y.~Constantinides, and G.~E.
  Karniadakis.
\newblock The flow dynamics of the garden-hose instability.
\newblock \emph{Journal of Fluid Mechanics}, 800:\penalty0 595--612, 2016.

\bibitem[Xu et~al.(2003)Xu, Rempfer, and Lumley]{xu2003turbulence}
S.~Xu, D.~Rempfer, and J.~Lumley.
\newblock Turbulence over a compliant surface: numerical simulation and
  analysis.
\newblock \emph{Journal of Fluid Mechanics}, 478:\penalty0 11--34, 2003.

\bibitem[Yeo(1990)]{yeo1990hydrodynamic}
K.~Yeo.
\newblock The hydrodynamic stability of boundary-layer flow over a class of
  anisotropic compliant walls.
\newblock \emph{Journal of Fluid Mechanics}, 220:\penalty0 125--160, 1990.

\end{thebibliography}

\bibliographystyle{abbrvnat}

\end{document}